\documentclass[11pt]{article}

\usepackage{amsmath}
\usepackage{graphicx}
\usepackage{indentfirst}
\usepackage{amssymb}
\usepackage{cite}
\usepackage{color}
\usepackage{subfigure}

\begin{document}

\title{\bf{Coalescence of Kerr Black Holes:\\
Binary Systems from GW150914 to GW170814}}

\date{}
\maketitle

\begin{center}
\author{Bogeun Gwak}$^a$\footnote{rasenis@sejong.ac.kr}\\

\vskip 0.25in
$^{a}$\it{Department of Physics and Astronomy, Sejong University, Seoul 05006, Republic of Korea}\\
\end{center}
\vskip 0.6in

{\abstract
{We investigate the energy of the gravitational wave from a binary black hole merger by the coalescence of two Kerr black holes with an orbital angular momentum. The coalescence is constructed to be consistent with particle absorption in the limit in which the primary black hole is sufficiently large compared with the secondary black hole. In this limit, we analytically obtain an effective gravitational spin--orbit interaction dependent on the alignments of the angular momenta. Then, binary systems with various parameters including equal masses are numerically analyzed. According to the numerical analysis, the energy of the gravitational wave still depends on the effective interactions, as expected from the analytical form. In particular, we ensure that the final black hole obtains a large portion of its spin angular momentum from the orbital angular momentum of the initial binary black hole. To estimate the angular momentum released by the gravitational wave in the actual binary black hole, we apply our results to observations at the Laser Interferometer Gravitational-Wave Observatory: GW150914, GW151226, GW170104, GW170608, and GW170814.
}
}

\thispagestyle{empty}
\newpage
\setcounter{page}{1}

\section{Introduction}\label{sec1}

The coalescence of black holes is one of the most important sources of gravitational waves. A gravitational wave occurs owing to a variation in the gravitational field, such as the motion of massive bodies. Although a gravitational wave can commonly occur owing to a small variation in the gravitational field, the magnitude of the gravitational wave in such a variation is too small to detect at our observable accuracy. Hence, an observable gravitational wave needs a sufficient magnitude so that it can be detected at observatories such as the Laser Interferometer Gravitational-Wave Observatory (LIGO). Since the magnitude of a gravitational wave becomes large along with the mass of its source, the candidate sources of detectable gravitational waves should be massive. Thus, the coalescence of black holes can be a source that releases a gravitational wave that is sufficiently large to be detected at the LIGO. Nowadays, several gravitational waves are detected at observatories. Most of their sources are the binary black hole mergers in GW150914, GW151226, GW170104, GW170608, GW170814\cite{TheLIGOScientific:2016src,Abbott:2016nmj,Abbott:2017vtc,Abbott:2017gyy,Abbott:2017oio}. Thus, the coalescence of black holes can be a frequently detectable source of gravitational waves. Further, a system of binary black holes has three angular momenta: the two spin angular momenta of the black holes and one orbital angular momentum, and these angular momenta play important roles in the analysis of the system. Note that the angular momenta for a binary black hole merger are described by the primary spin parameter $a_1$, secondary spin parameter $a_2$, and inspiral spin parameter $\chi_\text{eff}$ in the initial state and the spin parameter $a_\text{f}$ in the final state.

Black holes have conserved quantities such as the mass and angular momentum. There are two types of energies included with the mass of the black hole: a reducible energy and an irreducible mass. The roles of these two energies can be clarified in consideration of particle absorption to vary the black hole\cite{Christodoulou:1970wf,Bardeen:1970zz}. In particle absorption, the variations in a black hole¡¯s mass and angular momentum can be related to a particle's momenta such as the radial and angular momenta. Here, the black hole's angular momentum can be reduced by our choice of the particle's angular momentum; thus, part of the particle's angular momentum is a reducible energy in the variation of the black hole's mass. Interestingly, the remaining part cannot be reduced for any choice of the particle's momenta; therefore, the remaining part is called the irreducible mass of a black hole\cite{Christodoulou:1970wf,Christodoulou:1972kt}. This irreducible mass is physically defined as an energy distributed on the horizon of the black hole\cite{Smarr:1972kt}. For example, even if the mass of a black hole decreases during particle absorption, the irreducible mass still increases\cite{Christodoulou:1972kt}. Since this behavior of the irreducible mass is very similar to the entropy in thermodynamics, its square---the area of the horizon---is related to the Bekenstein--Hawking entropy\cite{Bekenstein:1973ur,Bekenstein:1974ax}. Further, the black hole can emit energy away from its horizon according to a quantum process. Owing to this small portion of the energy, the black hole can be assumed to be a thermal object having the Hawking temperature accounted from the emission\cite{Hawking:1974sw,Hawking:1976de}. For the variation in the black hole's mass, the Bekenstein--Hawking entropy and Hawking entropy act as thermodynamic variables in the first law of thermodynamics. In addition, the Bekenstein--Hawking entropy increases in an irreversible process in the second law of thermodynamics according to the behavior of the irreducible mass. Thus, the black hole itself can be considered as a thermal system obeying the laws of thermodynamics.

The black hole has a surface called the event horizon through which a particle or matter cannot escape from the gravity of the black hole. Then, the particle cannot be observed outside the horizon anymore. When absorbing a particle, the black hole also undergoes variations with respect to its physical properties. The stability of the black hole's horizon is one of its properties that is testable using particle absorption with respect to the black hole. The horizon of the black hole should be stable to cover the singularity located within it, because a naked singularity, an observable singularity without a horizon, causes a breakdown in the causality of spacetime. This is suggested as a weak cosmic censorship conjecture that prevents a naked singularity\cite{Penrose:1964wq,Penrose:1969pc}. Various investigations of the weak cosmic censorship conjecture have been applied to various black holes. The conjecture for a Kerr black hole was first tested by adding a particle\cite{Wald:1974ge}. Here, the conserved quantities carried by the particle cause a variation in the Kerr black hole with respect to its corresponding conserved quantities. By the addition of the particle's angular momentum, the angular momentum of the Kerr black hole can increase, but it cannot exceed the extremal limit of the black hole; thus, its horizon still exists and covers the singularity inside it. This implies that the weak cosmic censorship conjecture is valid when adding a particle. However, the validity of the conjecture depends on the test approach. For example, in a near-extremal Kerr black hole, the conjecture was found to be invalid\cite{Jacobson:2009kt}; thus, its angular momentum can be overspun owing to the addition of a particle. Then, the horizon disappears. This can be resolved by considering the self-force effect with the addition of a particle\cite{Barausse:2010ka,Barausse:2011vx,Colleoni:2015ena}. The conjecture for a Reissner--Nordstr\"{o}m (RN) black hole is valid when considering the back-reaction effect\cite{Hubeny:1998ga,Isoyama:2011ea}. There are also various tests of the conjecture for anti-de Sitter (AdS) black holes \cite{BouhmadiLopez:2010vc,Rocha:2011wp,Gao:2012ca,Rocha:2014gza,Rocha:2014jma,Colleoni:2015afa,McInnes:2015vga,Hod:2016hqx,Natario:2016bay,Horowitz:2016ezu,Duztas:2016xfg,Gwak:2017icn,Sorce:2017dst,Liang:2018wzd}. In particular, from a thermodynamic point of view, since the second law of thermodynamics ensures an increase in the area of the black hole's horizon, it can be a sufficient condition for the validity of the conjecture. This thermodynamic picture has been investigated for AdS black holes for particle absorption\cite{Gwak:2015fsa,Gwak:2016gwj,Gwak:2017kkt}.

The coalescence of black holes is also understood from a thermodynamic point of view. As a thermal system, the coalescence can be an irreversible process in which the initial two black holes become a final black hole given a thermal preference. The increase in the entropy due to the irreversible process, the second law of thermodynamics, was first shown for the coalescence of two Schwarzschild black holes\cite{Hawking:1971tu}. Further, during coalescence, the energy released by the gravitational wave is expected to be in terms of the upper limit of the energy of the gravitational wave under the second law of thermodynamics. However, the energy limit is much larger than the real energy of the gravitational wave; therefore, there are some difficulties in deriving detailed physical implications from the limit. In spite of these difficulties, when applied to the coalescence of Kerr black holes, the upper limit indicates the existence of an effective gravitational spin interaction between black holes\cite{Wald:1972sz} where attraction or repulsion acts on the black holes according to their alignments. For the case where one of black holes is much smaller than the other black hole, the form of the interaction is exactly coincident with the interaction potential acting on a particle spinning around the Kerr black hole, as already obtained from the Mathisson--Papapetrou--Dixon (MPD) equations\cite{Mathisson:1937zz,Papapetrou:1951pa,Dixon:1970zza,Schiff:1960gi,Mashhoon:1971nm,Wilkins:1970wap,Majar:2012fa,Zilhao:2013nda,Plyatsko:2015bia,d'Ambrosi:2015xci}. Hence, a gravitational spin interaction is induced from the coupling of the spin angular momenta of the two black holes. When the spin angular momentum of a black hole is sufficiently large, as in Myers--Perry (MP) black holes where there is no extremal limit for the spin angular momentum in higher-dimensional spacetime, the interaction between MP black holes plays an important role in the instability of the black hole system and the upper limit of the energy released by the gravitational wave\cite{Gwak:2016cbq}. Further, using the upper limit, the various constants of a given model of a regular black hole can be specified with physically possible ranges by matching LIGO observations\cite{Gwak:2017zwm}.

In this paper, we will investigate the gravitational wave released from a binary black hole merger by the coalescence of two Kerr black holes with an orbital angular momentum. Since an astrophysical black hole can be theoretically approximated as a Kerr black hole\cite{TheLIGOScientific:2016src,Yunes:2016jcc}, we consider a binary system consisting of two spin angular momenta due to Kerr black holes and one orbital angular momentum for their orbit. In particular, the orbital angular momentum is found to be important to the state of the final black hole because its angular momentum is provided by the orbital angular momentum. During coalescence, we assume conservation of the system's energy and angular momentum. In addition, the irreducible mass of Kerr black holes is assumed to be irreducible. Since the irreducible mass cannot be extracted by a physical process, even a Penrose process\cite{Bardeen:1970zz,Penrose:1971uk}, it can be expected to not decrease during the coalescence of black holes, an irreversible process. This implies that the irreducible mass is used for the formation of the final black holes, and a gravitational wave is released from the reducible mass such as the kinetic and rotational energies in the initial state. Our assumptions for coalescence are consistent with the second law of thermodynamics. In the particle limit for one of the black holes, our model is proven to be consistent with particle absorption\cite{Christodoulou:1970wf,Christodoulou:1972kt} and the MPD equations\cite{Wald:1972sz}. Further, we analytically obtain that the energy released by the gravitational wave depends on the effective gravitational spin--spin and spin--orbit interactions. In particular, the spin--orbit interaction is newly obtained from an effective force and is one of the advantages of our model, which considers the orbital angular momentum. Then, we numerically compute the final black hole and the energy released by the gravitational wave to obtain the effects of various variables during coalescence for the equal-mass case. Interestingly, the obtained energy of the gravitational wave is within a very similar range as those of LIGO observations\cite{TheLIGOScientific:2016src,Abbott:2016nmj,Abbott:2017vtc,Abbott:2017gyy,Abbott:2017oio}; therefore, we can ensure that our results are relevant to the understanding of an actual binary black hole merger. This is another advantage of our model. By numerical computation, the relations between the alignments of the black holes and the gravitational wave are found to be consistent with the effective interactions. In addition, the orbital angular momentum also shows similar behaviors to the spin angular momentum in its effect. Finally, we apply our results to five recent LIGO observations to find the relevant range of angular momenta in our model. Here, we obtain the initial spin parameters, orbital angular momentum, total angular momentum, and angular momentum of the gravitational wave. Their median values are set to be very consistent with the conservation of angular momentum. In particular, the orbital angular momentum is the largest portion of the total angular momentum in the initial state. Hence, it can be important to the formation of a black hole with a high spin parameter during coalescence.

This paper is organized as follows. In section\,\ref{sec2}, we review Kerr black holes. Particle absorption and the MPD equations are also introduced to show the analytical forms of the effective interactions. In section\,\ref{section3}, we construct the basic framework of our model. Then, to prove its consistency with particle absorption and the MPD equations, we compute a variation of our model in the particle limit. In this process, we will prove that the effective interactions predicted by our model are coincident with those of particle absorption and the MPD equations. In section\,\ref{section4}, we numerically obtain the final states in equal-mass cases. Here, the effective interactions are obtained as analytical forms and found to work as predicted in the particle limit. In section\,\ref{section5}, we apply our model to recent LIGO observations of binary black hole mergers. We compute initial spin parameters by using other parameters. Further, the orbital angular momentum and the angular momentum of the released gravitational wave are obtained by our model. In section\,\ref{sec6}, we briefly summarize our results.

\section{Thermodynamics and Spin Interaction of Kerr Black Hole}\label{sec2}

We will assume that the binary black hole is a system consisting of two Kerr black holes with an orbital angular momentum in their orbit. The angular momenta of the binary black hole play important roles in the effective interactions between black holes. Further, during the coalescence of the binary black hole, the energy released by the gravitational wave is significantly affected by the interactions. In this work, we will construct the coalescence of the binary black hole that satisfies the laws of thermodynamics. In this procedure, the effective interactions related to the angular momenta of the system will be shown to be consistent with what are expected from the particle absorption and MPD equations at the particle limit of one of the black holes. Hence, we will review the basics of Kerr black holes, particle absorption, and the gravitational spin interaction from the MPD equations.

\subsection{Basics of Kerr Black Hole}

The Kerr black hole is a solution to the Einstein equations in four-dimensional spacetime. The metric of a Kerr black hole with a mass $M$ and spin angular momentum $J$ in Boyer--Lindquist coordinates is
\begin{align}\label{eq:Kerr}
ds^2&=-\frac{\Delta}{\rho^2}\left(dt-{a\sin^2\theta} d\phi\right)^2+\frac{\rho^2}{\Delta}dr^2+{\rho^2}d\theta^2+\frac{\sin^2\theta}{\rho^2}\left(a dt-(r^2+a^2)d\phi\right)^2,\\
\rho^2&=r^2+a^2\cos^2\theta,\quad\Delta=(r^2+a^2)-2Mr,\quad J=Ma,\nonumber
\end{align}
where the spin parameter is given as $a$, which is bounded when $M\geq a$. For $M>a$, the Kerr black hole has two event horizons.
\begin{align}
r_\text{in} = M-\sqrt{M^2-a^2}, \quad r_\text{h} = M+\sqrt{M^2-a^2}, 
\end{align}
where $r_\text{in}$ and $r_\text{h}$ denote the inner and outer horizons, respectively. Since the inside of the outer horizon cannot be seen by an asymptotic observer, the properties of the Kerr black hole are defined at its outer horizon. An asymptotic observer measures the angular velocity of spacetime at the outer horizon as
\begin{align}
\Omega_\text{h}=\frac{a}{r_\text{h}^2 +a^2}.
\end{align}
The Hawking temperature and Bekenstein--Hawking entropy are
\begin{align}\label{eq:definitions07}
T_\text{h}=\frac{r_h\left(1-\frac{a^2}{r_\text{h}^2}\right)}{4\pi (r_\text{h}^2+a^2)}, \quad S_\text{h} = \pi (r_\text{h}^2+a^2).
\end{align}
For a given mass, the spin parameter $a$ is saturated at $M=a$, where the Kerr black hole satisfies the extremal condition. Under the extremal condition, the inner and outer horizons are coincident with each other. Then,
\begin{align}
r_\text{h}=r_\text{in}=M.
\end{align}
When the spin parameter exceeds the value of the mass, $M<a$, there is no horizon covering the inside of the black hole. Then, the curvature singularity of spacetime is exposed to the observer located outside. This is called a naked singularity. However, according to the cosmic censorship conjecture, it is expected that there is no physical process for overspinning a Kerr black hole into a naked singularity\cite{Penrose:1964wq,Penrose:1969pc}. This can be shown by particle absorption, among the various verifications investigated for the cosmic censorship conjecture for Kerr black holes\cite{Wald:1974ge}.

\subsection{Thermodynamics of Kerr Black Hole under Particle Absorption}

In particle absorption, the acceleration of a Kerr black hole's angular velocity is considered by adding a particle. Owing to the energy and angular momentum of the particle, the corresponding conserved quantities of the Kerr black hole vary in the energy equation of the particle. In these variations, the mass of the Kerr black hole is divided into two parts: the reducible energy and irreducible mass\cite{Christodoulou:1970wf,Christodoulou:1972kt}. In a Kerr black hole, the reducible energy includes the rotational and kinetic energies. Further, the irreducible mass is a type of energy distributed on the surface of the horizon\cite{Smarr:1972kt}. We will introduce the irreducible mass by using particle absorption. To obtain a relation between the conserved quantities of a particle passing through the outer horizon of the Kerr black hole, the particle's equations of motion are obtained using Hamilton--Jacobi method.
\begin{align}
\mathcal{H}=\frac{1}{2}g^{\mu\nu} p_\mu p_\nu,\quad S=\frac{1}{2}m^2\lambda-Et+L\phi+S_r(r)+S_\theta(\theta),
\end{align}
which are the Hamiltonian and Hamilton--Jacobi action of a particle, respectively, in the metric of a Kerr black hole. Using a separate variable $\mathcal{K}$\cite{Carter}, we can write the radial and $\theta$-direction equations of motion as
\begin{align}\label{eq:twoeqs}
\frac{\partial r}{\partial \lambda}\equiv\dot{r}\equiv p^r &= \frac{\Delta \sqrt{R(r)}}{\rho^2},\quad \frac{\partial \theta}{\partial \lambda}\equiv\dot{\theta}\equiv p^\theta = \frac{\sqrt{\Theta(\theta)}}{\rho^2},
\end{align}
with
\begin{align}
\partial_r S_r(r)&\equiv \sqrt{R(r)}\equiv\sqrt{\frac{1}{\Delta^2}\left(aL-(r^2+a^2)E\right)^2-\frac{m^2 r^2+\mathcal{K}}{\Delta}},\nonumber\\
\partial_\theta S_\theta(\theta)&\equiv \sqrt{\Theta(\theta)}\equiv\sqrt{\mathcal{K}-a^2 m^2 \cos^2\theta-\left(L\csc\theta -a E \sin\theta \right)^2}.\nonumber
\end{align}
By combination with Eq.\,(\ref{eq:twoeqs}), the removal of $\mathcal{K}$ gives the particle's energy equations for a given location. Then,
\begin{align}\label{eq:energyeqs02}
\alpha E^2 +\beta E +\gamma =0,
\end{align}
where
\begin{align}
\alpha&=\frac{(r^2+a^2)^2}{\Delta}-a^2 \sin^2\theta,\quad \beta=-\frac{2aL(r^2+a^2-\Delta)}{\Delta}\nonumber\\
\gamma&=-\frac{1}{\Delta}\left(-a^2 L^2 +m^2 r^2 \Delta +((p^r)^2 + (p^\theta)^2\Delta)\rho^4+a^2m^2 \Delta \cos^2\theta +L^2\Delta \csc^2\theta\right).\nonumber
\end{align}
The particle is assumed to be absorbed into the Kerr black hole when it passes through its outer horizon. At that moment, the relation between the energy and momenta of the particle is given by the energy equation in Eq.\,(\ref{eq:energyeqs02}) which becomes
\begin{align}
((r_h^2+a^2)E-aL )^2=\rho_h^4 (p^r)^2,
\end{align}
at the outer horizon. Its solution is\cite{Christodoulou:1970wf,Christodoulou:1972kt}
\begin{align}\label{eq:relation01}
E=\frac{aL}{r_h^2+a^2} + \frac{\rho_h^2}{r_h^2+a^2}|p^r|,
\end{align}
where we choose the $(+)$ sign among the two solutions because the particle is coming into the Kerr black hole in a positive time flow. Then, the solution shows the relation between the particle's energy and momenta at the outer horizon. When the particle passes through the outer horizon, the energy and angular momentum of the particle are assumed to be those of the Kerr black hole. Then, the conserved quantities of the black hole vary as much as those of the particle; thus, 
\begin{align}\label{eq:particleabsorption02}
dM = E,\quad dJ = L.
\end{align}
For particle absorption, we can write the relation between the variations in the mass and angular momentum of the Kerr black hole using Eq.\,(\ref{eq:relation01}). Then,
\begin{align}\label{eq:relation02}
dM=\frac{a}{r_h^2+a^2}dJ + \frac{\rho_h^2}{r_h^2+a^2}|p^r|,
\end{align}
which constrains the variation in the Kerr black hole in the particle absorption process\cite{Wald:1974ge}. Under this constraint, the variation in the entropy of the Kerr black hole becomes
\begin{align}\label{eq:varentropy}
dS_\text{h}=d\left(\pi(r_\text{h}^2+a^2)\right)=\frac{2\pi\rho_\text{h}^2}{ (r_\text{h}-M)}|p^r|,
\end{align}
which is always positive because $r_\text{h}>M$, and the equality $r_\text{h}=M$ also gives an increase in the entropy or surface of the horizon. This is consistent with the second law of thermodynamics\cite{Gwak:2015fsa}. We can obtain the first law of thermodynamics by inserting Eq.\,(\ref{eq:varentropy}) into Eq.\,(\ref{eq:relation02})\cite{Gwak:2017kkt}. Then,
\begin{align}
dM = T_\text{h}dS_\text{h}+\Omega_\text{h} dJ.
\end{align}
Therefore, particle absorption varies the Kerr black hole and satisfies the laws of thermodynamics. Here, we can obtain an interesting property of black holes for particle absorption. Rewritten Eq.\,(\ref{eq:relation02}) becomes an inequality as  
\begin{align}\label{eq:relation03}
dM-\frac{a}{r_h^2+a^2}dJ = \frac{\rho_h^2}{r_h^2+a^2}|p^r|\geq 0,
\end{align}
where the left-hand side is an irreducible property in the process. By integrating out the left-hand side of Eq.\,(\ref{eq:relation03}), we can define a property having the same dimension as the mass\cite{Christodoulou:1970wf,Christodoulou:1972kt}.
\begin{align}
M_\text{ir}=\frac{1}{2}\sqrt{r_\text{h}^2+a^2},
\end{align}
which is called the irreducible mass\cite{Christodoulou:1970wf,Christodoulou:1972kt,Bekenstein:1973ur}. The irreducible mass is assumed to be an energy distributed on the surface of the horizon\cite{Smarr:1972kt}. Then, in terms of the irreducible mass, the mass of the Kerr black hole can be divided into the irreducible mass and a rotation energy such that\cite{Misner:MTW}
\begin{align}\label{eq:irreduciblemass1}
M=M(M_\text{ir},J)=\sqrt{M_\text{ir}^2+\frac{J^2}{4M_\text{ir}^2}}.
\end{align}
Hence, the mass of the Kerr black hole actually consists of irreducible and reducible masses. Then, by a physical process such as a Penrose process\cite{Penrose:1971uk}, the irreducible mass still increases in a physical process, even if the mass of the Kerr black hole can be reduced owing to the extraction of the rotational energy in Eq.\,(\ref{eq:irreduciblemass1}).

\subsection{Gravitational Spin--Spin Interaction in MPD Equations}

We will consider a binary black hole system; thus, the binary system can expected to have an effective gravitational spin interaction between two black holes with their spin angular momenta. Using the second law of thermodynamics, the increase of the entropy, the contribution of the spin--spin interaction is estimated in the energy of the gravitational wave in the coalescence of the black holes, and the form of the spin--spin interaction potential derived from the MPD equations for a spinning particle is clearly coincident in the limit where one of black holes is much smaller and slowly rotating compared with the other black hole\cite{Wald:1972sz}. In this section, we review these results with an introduction to the MPD equations\cite{Mathisson:1937zz,Papapetrou:1951pa,Dixon:1970zza} in the Kerr black hole spacetime. We consider a Kerr black hole with a mass $M_1$ and an angular momentum $J_1=M_1 a_1$ and a spinning particle with a mass $M_2$ and an angular momentum $J_2$. The spinning particle is also has a four-velocity $v^\mu$ and linear momentum $p^a$. In the MPD equations, the momentum and four-velocity have a difference related to a proper time $s$ due to the spinning effect. Then,
\begin{align}\label{eq:mpdeqs02}
\frac{D p^a}{D s} = -\frac{1}{2}R^a_{bcd} v^b S^{cd},\quad \frac{D S^{ab}}{D s} = p^a v^b - p^b v^a, \quad S_a = \frac{1}{2 M_2} \sqrt{-g} \epsilon_{abcd} p^b S^{cd},
\end{align}
where $R^a_{bcd}$ is the Riemann curvature tensor of the Kerr metric. The spin of the particle is given in terms of the spin tensor $S^{ab}$ and spin vector $S_a$. The trajectory of the spinning particle can be derived by imposing the supplementary condition\cite{Beiglbock:1967fun}
\begin{align}\label{eq:supplementary01}
p_a S^{ab}=0.
\end{align}
Then, we can determine the motion of the spinning particle using Eqs.\,(\ref{eq:mpdeqs02}) and (\ref{eq:supplementary01}). The mass and spin angular momentum of the spinning particle are defined as
\begin{align}
J_2^2 = \frac{1}{2}S_{ab}S^{ab},\quad M_2^2 = -p_a p^a,\quad p^a = M_2 v^a.
\end{align} 
To obtain the form of the spin interaction potential, we assume that the axes of the spin angular momenta of both the black hole and particle are coincident with each other for simplicity and that the particle slowly comes into the black hole, such that the particle is assumed to be nonrelativistic, $v^a\ll 1$. Then, the initial state for the velocity $v^a$ and spin vector $S^a$ of the particle is given as
\begin{align}
v^a = \left(\frac{1}{\sqrt{-g_{tt}}},v^a,0,0\right),\quad J^a_2 = \left(0,\frac{J_2}{\sqrt{g_{rr}}},v^a,0,0\right),
\end{align}
where the two vectors are normalized. The interaction potential between the black hole and the particle is obtained from the energy of the spinning particle, which is a conserved quantity with respect to the Killing vector $\xi^t$ in terms of the time coordinate. Hence, the energy of the spinning particle is derived as
\begin{align}\label{eq:energyspinning01}
E=-p_t - \frac{1}{2}S^{ab} \nabla_a g_{bt},
\end{align}
where the first term is related to the kinetic energy, and the second term is the spin interaction with respect to the spin tensor. Then, the second term in Eq.\,(\ref{eq:energyspinning01}) becomes
\begin{align}\label{eq:spininteraction05}
U_\text{spin,int}=\frac{J_1 J_2}{M_1(r_1^2+a_1^2)},
\end{align}
which is exactly coincident with the spin interaction derived from the second law of thermodynamics\cite{Wald:1972sz}. The sign of the spin potential in Eq.\,(\ref{eq:spininteraction05}) depends on the alignment between $J_1$ and $J_2$ and implies effective attraction and repulsion between the black hole and the particle. The parallel alignment, a plus sign, has a positive potential; thus, the effective force acts in a repulsive manner. The antiparallel alignment has a negative sign; hence, the effective force acts in an attraction. This effective interaction plays an important role in the gravitational wave released in the collision of the black holes. We will derive using our approach in the particle limit and investigate the effect of the potential in following sections. Note that we will use dimensionless coordinates and variables scaled by the solar mass $M_\odot$, such as
\begin{align}
\tilde{r}=\frac{r}{M_\odot},\quad\tilde{M}=\frac{M}{M_\odot},\quad \tilde{M}_\text{ir}=\frac{M_\text{ir}}{M_\odot},\quad \tilde{a}=\frac{a}{M_\odot}, \quad \tilde{J}=\frac{J}{M_\odot^2},
\end{align}
where we will omit the tildes for convenience.

\section{Basic Framework}\label{section3}

The effects of the angular momenta in a binary black hole merger as a source of the gravitational wave detected at the LIGO will be investigated. Here, we assume an initial binary black hole in a model that consists of two Kerr black holes with an orbital angular momentum. Then, the binary black hole merger produces a final Kerr black hole with the released gravitational wave. In this model, the energy of the gravitational wave can be estimated, satisfying the laws of thermodynamics between the initial and final states. In the initial state, two Kerr black holes are located far from each other; thus, their gravitational interaction can be ignored. These Kerr black holes rotate with the orbital angular momentum $L_\text{orb}$, which will be included in the total angular momentum. The primary and secondary black holes are $(M_1,a_1)$ and $(M_2,a_2)$ in the initial state, and their axes of spin angular momenta have an angular difference $\psi$. These angular momenta will play an important role in explaining the final state of the black hole system. According to Eq.\,(\ref{eq:irreduciblemass1}), the energies of the initial state are divided into an irreducible mass and a rotation energy including the orbital angular momentum. Then, as they slowly come together with a spiral motion due to the orbital angular momentum, the two Kerr black holes merge into a Kerr black hole of $(M_\text{f},a_\text{f})$ in the final state. Since the total energy of the system should be conserved in the coalescence, the released gravitational wave is equivalent to the loss of mass between the initial and final states. Then, the energy and angular momentum of the gravitational wave are
\begin{align}\label{eq:finalstate02}
M_\text{gw}=(M_\text{1}+M_\text{2})-M_\text{f},\quad \vec{J}_\text{gw}=(\vec{J}_\text{1}+\vec{J}_\text{2}+\vec{L}_\text{orb})-\vec{J}_\text{f},
\end{align}
where, for simplicity, we assume in the initial state that the sum of $\vec{J}_1$ and $\vec{J}_2$ is aligned with $\vec{L}_\text{orb}$; thus, $\vec{J}_\text{f}$ and $\vec{J}_\text{gw}$. Then, the magnitude of the sum of angular momenta in the initial state becomes
\begin{align}\label{eq:alignmentangle015}
|\vec{J}_\text{tot}|=|\vec{J}_\text{1}+\vec{J}_\text{2}+\vec{L}_\text{orb}|=\sqrt{J_\text{1}^2+J_\text{2}^2+2 J_\text{1}J_\text{2}\cos\psi}+L_\text{orb},
\end{align}
The ratios of the mass and angular momentum of the gravitational wave with respect to the total mass and angular momentum in the initial state are respectively defined as
\begin{align}\label{eq:ratios01}
\epsilon_\text{M} = \frac{M_\text{gw}}{M_\text{1}+M_\text{2}},\quad \epsilon_\text{J} = \frac{\vec{J}_\text{gw}}{|\vec{J}_\text{tot}|}.
\end{align}
To estimate the value of $M_\text{gw}$, we assume that the total irreducible mass increases during coalescence. Since the irreducible mass cannot be extracted or decreased by the Penrose process in the Kerr black hole, our assumption can be a reasonable generalization of particle absorption for the coalescence of black holes. This is our main assumption in this work. Fortunately, from a thermodynamic point of view, this will provide quite precise predictions about the energy of the gravitational wave released in the coalescence compared with its upper limits in \cite{Hawking:1971tu,Wald:1972sz,Gwak:2016cbq,Gwak:2017zwm}. Then, the increase in the irreducible mass from the initial state to the final state is
\begin{align}\label{eq:increaseirreduciblemass}
M_\text{1,ir}+M_\text{2,ir} \leq M_\text{f,ir}.
\end{align}
Our assumption in Eq.\,(\ref{eq:increaseirreduciblemass}) satisfies the second law of thermodynamics. According to the definition of the Bekenstein--Hawking entropy in Eq.\,(\ref{eq:definitions07}), the increase in the irreducible mass in Eq.\,(\ref{eq:increaseirreduciblemass}) becomes a sufficient condition; hence,
\begin{align}
S_\text{bh,i1}+S_\text{bh,i2}< S_\text{bh,f}.
\end{align}
Therefore, the entropy of the system increases in the process of coalescence. Then, our assumption is relevant to the second law of thermodynamics. Physically, the increase in the irreducible mass implies that a Kerr black hole is approximated as a solid body with a spin angular momentum. Hence, most of the energy of the gravitational wave is released from the reducible energy $M_\text{re}$ included in the mass $M$. Here, the reducible energy is simply assumed to the difference between the mass and irreducible mass of the Kerr black hole. Owing to our assumption for the initial condition, the main part of the reducible energy is the rotational energy, so we define the reducible energy by the rotational energy: $M_\text{rot}=M-M_\text{ir}$. In combination with Eqs.\,(\ref{eq:finalstate02}), (\ref{eq:ratios01}), and (\ref{eq:increaseirreduciblemass}), the upper limit of the energy of the gravitational wave can be estimated as
\begin{align}
M_\text{gw}\leq M_\text{gw,upper}.
\end{align}
The upper limit of the energy of the gravitational wave precisely approaches the real value of the gravitational wave in LIGO observations with our model. It will be investigated in following sections. There are two ratios related to the mass and angular momentum. We will find constraints on $\epsilon_\text{M}$ and $\epsilon_\text{J}$ considering particle absorption. Further, we will obtain a type of effective interaction related to the orbital angular momentum $L_\text{orb}$.

\subsection{Ratios Related to the Mass and Angular Momentum}

Our assumption for the coalescence of Kerr black holes should be consistent with particle absorption when $M_1 \gg M_2$, where the primary black hole is fixed as a background and the secondary black hole is treated as a particle. Hence, physical constraints on $\epsilon_\text{M}$ and $\epsilon_\text{J}$ can be obtained from particle absorption. In particle absorption, the ratios $\epsilon_\text{M}$ and $\epsilon_\text{J}$ become very small because $M_1\gg M_2$ and $M_\text{gw}\ll 1$, where we will assume that $M_1=M$ and $M_2=E$. The angular momentum corresponds to the orbital angular momentum; thus, $L_\text{orb}=L$, and $J_1\gg L$ because the particle has only an orbital angular momentum. Then,
\begin{align}\label{eq:gwupper01}
\epsilon_\text{M} = \frac{M_\text{gw}}{M+E},\quad \epsilon_\text{J} = \frac{J_\text{gw}}{J+L},
\end{align}
and from Eq.\,(\ref{eq:finalstate02}),
\begin{align}\label{eq:gwupper02}
M_\text{gw}=(M+E)-(M+dM),\quad J_\text{gw}=(J+L)-(J+dJ).
\end{align}
By a combination of Eqs.\,(\ref{eq:gwupper01}) and (\ref{eq:gwupper02}),
\begin{align}\label{eq:epsilons02}
(M+E)(1-\epsilon_\text{M})=M+dM,\quad (J+L)(1-\epsilon_\text{J})=J+dJ.
\end{align}
Since the primary black hole is the given background and the variables are the particle's energy and momenta, we have to assume that $(1-\epsilon_\text{M})M\approx M$ and $(1-\epsilon_\text{J})J\approx J$ in particle absorption. This implies that there is no radiation from the primary black hole as the background. Then, all of the variations in the black hole originate from the variables of the particle. The coalescence of black holes is not technically coincident with particle absorption; hence, we need to modify this part. Then, the particle contributes the variation in the primary black hole as
\begin{align}\label{eq:particleabsorpton03}
dM\approx (1-\epsilon_\text{M})E,\quad dJ\approx (1-\epsilon_\text{J})L,
\end{align}
which imply that part of the particle's energy is absorbed into the black hole. The other part is released by the gravitational wave. This is also consistent with Eq.\,(\ref{eq:particleabsorption02}) with addition of ratios. Using Eq.\,(\ref{eq:particleabsorpton03}), the relation in Eq.\,(\ref{eq:relation01}) becomes
\begin{align}\label{eq:relation03}
\frac{dM}{1-\epsilon_\text{M}}=\frac{a}{(1-\epsilon_\text{J})(r_\text{h}^2+a^2)}dJ + \frac{\rho_\text{h}^2}{r_\text{h}^2+a^2}|p^r|,
\end{align}
from which the variation in the entropy becomes
\begin{align}\label{eq:epparticleabsorption02}
dS_\text{h}=\frac{2\pi (\epsilon_\text{J}-\epsilon_\text{M})}{r_\text{h}-M}L+\frac{2\pi (1-e_\text{M})\rho_\text{h}^2}{r_\text{h}-M}|p^r|.
\end{align} 
Owing to the second law of thermodynamics, the entropy should increase in particle absorption owing to the irreversible process. The sign of the first term in Eq.\,(\ref{eq:epparticleabsorption02}) depends on the alignment between $J$ and $L$; thus, we assume that $\epsilon_\text{M}\approx \epsilon_\text{J}$ to remove the dependency on the alignment. Then, the entropy becomes irreducible. Further, this assumption is reasonable because the angular momentum of a black hole is proportional to its mass, i.e., $J\sim M$; thus, the ratio of emitted angular momentum will be as much as that of emitted mass between the initial and final states. This ensures the increase in the irreducible mass of the system under $\epsilon_\text{M}=\epsilon_\text{J}$ as
\begin{align}
dM_\text{ir}=\frac{(1-\epsilon_\text{M})\rho_\text{h}^2 |p^r|}{2(r_\text{h}-M)\sqrt{r_\text{h}^2+a^2}}.
\end{align}
Here, the initial condition related to the slow approach of the black holes during their coalescence can be converted to $|p^r|\approx 0$ in particle absorption. Therefore, we will assume that 
\begin{align}\label{eq:assumeeqs015}
\epsilon_\text{M}\approx \epsilon_\text{J},\quad dM_\text{ir}\approx 0.
\end{align}
Note that our assumptions just introduce minor effects on the gravitational wave, but this reduces the number of variables related to ratios to one.

\subsection{Gravitational Spin--Orbit Interaction from Particle Absorption}

Since most of the binary black holes observed at the LIGO are expected to have antiparallel alignment, as implied by the inspiral spin parameters $\chi_\text{eff}$ approximately having values of zero, instead of the black hole's spin angular momenta canceling each other, the orbital angular momentum of the binary black hole plays an important role in the spin angular momentum of the final black hole. Here, we will investigate the effects of the orbital angular momentum on the configuration of the final black hole and the gravitational wave in the case where $M_1\gg M_2$ and $L_\text{orb}\ll M_1^2$, consistent with particle absorption. The secondary black hole is assumed to have a spin angular momentum of zero to show the contribution of the orbital angular momentum clearly. The variation in the orbital angular momentum in the initial state affects the final black hole and gravitational wave; hence, from Eq.\,(\ref{eq:finalstate02}),
\begin{align}\label{eq:vargravit03}
\frac{\partial M_\text{gw}}{\partial L_\text{orb}}=-\frac{\partial M_\text{f}}{\partial L_\text{orb}}.
\end{align}
Then, we impose an irreducible mass for the equality in Eq.\,(\ref{eq:increaseirreduciblemass}) because $p^r\approx 0$, and its variation becomes
\begin{align}\label{eq:varirredu04}
a_\text{f} \frac{\partial a_\text{f}}{L_\text{orb}}+r_\text{f} \frac{\partial r_\text{f}}{L_\text{orb}}=0, \quad r_\text{f}=M_\text{f}+\sqrt{M_\text{f}^2+a_\text{f}^2}.
\end{align}
By combining Eqs.\,(\ref{eq:vargravit03}) and (\ref{eq:varirredu04}), the variation in the final black hole's mass with respect to the orbital angular momentum is obtained as
\begin{align}
\frac{\partial M_\text{f}}{\partial L_\text{orb}}=\frac{a_\text{f}}{r_\text{f}^2+a_\text{f}^2},
\end{align}
which is exactly coincident with the contribution of the particle's angular momentum for particle absorption in Eq.\,(\ref{eq:relation02}). Thus, the mass of the final black hole affected by the orbital angular momentum is obtained as
\begin{align}
M_\text{f}=\frac{J_1}{M_\text{f} (r_\text{1}^2+a_\text{1}^2)}L_\text{orb},
\end{align}
where we impose that $M_\text{f}\approx M_1$, and $a_\text{f}\approx a_1$. Owing to Eq.\,(\ref{eq:vargravit03}), the change in the final black hole's mass is opposite of the energy of the released gravitational wave with respect to the orbital angular momentum. Then, the orbital angular momentum contributes to the energy of the gravitational wave by
\begin{align}\label{eq:intpoten05}
M_\text{gw}= -\frac{J_1}{M_1(r_\text{1}^2+a_\text{1}^2)}L_\text{orb}.
\end{align}
The effect of the orbital angular momentum can be considered as an effective interaction potential of the orbital angular momentum in a black hole system. The accumulated energy for the potential is released in terms of the gravitational wave. Hence, the sign of the interaction potential is opposite to that of $M_\text{gw}$; hence,
\begin{align}\label{eq:intpoten02}
U_\text{orb,int}=\frac{J_1}{M_1(r_\text{1}^2+a_\text{1}^2)}L_\text{orb}.
\end{align}
This is the spin--orbit interaction potential which can be obtained in our model having the orbital angular momentum. In the coalescence of a binary black hole, the energy of the gravitational wave contributes as much as the interaction potential in Eq.\,(\ref{eq:intpoten02}). Then, the antiparallel alignment between the primary black hole's angular momentum and the orbital angular momentum releases more energy of the gravitational wave than their parallel alignment owing to the contribution of the interaction potential because the interaction potential is $U_\text{orb,int}<0$ in the antiparallel case and $U_\text{orb,int}>0$ in the parallel case. This implies that the antiparallel alignment releases potential energy owing to the attraction, but the parallel alignment needs to overcome its repulsion; thus, less energy is released compared with the antiparallel case. Note that attraction or repulsion can be easily shown by the sign of the interaction potential in Eq.\,(\ref{eq:intpoten02}). Effectively, the energy released by the gravitational wave is equal to the interaction potential.

\subsection{Gravitational Spin--Spin Interaction from MPD Equations}

Here, we will investigate the effects of the spin parameter of the black hole. This will show that the alignment between black holes also plays an important role in the emitted gravitational wave. The procedure is similar to that in the previous section. For the collision of two black holes, in which one of them is a slowly rotating black hole with a small mass, $M_2\ll M_1$, and $a_2\ll M_1$ in the initial state. The gravitational wave with respect to the variation in $a_2$ becomes
\begin{align}\label{eq:relations031}
\frac{\partial M_\text{gw}}{\partial a_2}=-\frac{\partial M_\text{f}}{\partial a_2}.
\end{align}
The variation in Eq.\,(\ref{eq:increaseirreduciblemass}) with respect to $a_2$ is
\begin{align}
\frac{1}{\sqrt{r_\text{f}^2+a_\text{f}^2}}\left(r_\text{f}\frac{\partial r_\text{f}}{\partial a_2}+a_\text{f} \frac{\partial a_\text{f}}{\partial a_2}\right)=\frac{1}{\sqrt{r_2^2+a_2^2}}\left(r_2 \frac{\partial r_2}{\partial a_2}+a_2\right).
\end{align}
Then, the variation in the mass of the final black hole with respect to $a_2$ is obtained as
\begin{align}\label{eq:spininteraction015}
\frac{\partial M_\text{f}}{\partial a_2} = \frac{M_2 a_1}{r_1^2+a_1^2}\,,
\end{align}
which implies that the mass of the final black hole is partially affected by the spin parameter $a_2$. By integrating Eq.\,(\ref{eq:spininteraction015}) with Eq.\,(\ref{eq:relations031}), the energy of the gravitational wave is partially emitted from the energy depending on spin angular momenta, which is
\begin{align}\label{eq:spininteraction012}
M_\text{gw} = -\frac{J_1 J_2}{M_1(r_1^2+a_1^2)}\,,
\end{align}
The spin interaction can be written in terms of the spin interaction potential as
\begin{align}\label{eq:gintpot05}
U_\text{spin,int}=\frac{J_1 J_2}{M_1(r_1^2+a_1^2)}.
\end{align}
This is exactly coincident with the gravitational interaction potential predicted by the MPD equations in Eq.\,(\ref{eq:spininteraction05}). Thus, our approach for the binary black hole is consistent with particle absorption and the MPD equations in the particle limit of the secondary black hole. The spin interaction potential in Eq.\,(\ref{eq:gintpot05}) changes its sign with $J_1$ and $J_2$. The interaction is attractive for $J_1 J_2 <0$ or repulsive for $J_1 J_2 >0$. Then, the antiparallel alignment releases more energy than the parallel alignment during the coalescence of the binary black hole. Note that our analysis based on the irreducible mass provides the same result as the MPD equations and entropy-based analysis provided in \cite{Wald:1972sz}.

\section{Energy of Gravitational Wave in Coalescence of Kerr Black Holes}\label{section4}

We will numerically investigate the gravitational wave released in the collision of two Kerr black holes when $M_\text{1}\approx M_\text{2}$. Since our approach from Eqs.\,(\ref{eq:finalstate02})--(\ref{eq:increaseirreduciblemass}) is consistent with the particle limits when $M_\text{1}\gg M_\text{2}$, we will apply our model to various black hole pairs to investigate the energy of the gravitational wave. Further, the effects of variables such as the spin and orbital angular momenta will be studied. Once again, we assume the coalescence of two Kerr black holes such that
\begin{align}
\epsilon=\epsilon_\text{M}\approx\epsilon_\text{J}, \quad |p^r|\approx 0,
\end{align}
where the second equation related to the radial momentum means that the black holes slowly come together in the radial direction; therefore, the irreducible mass of the system is almost conserved, as shown in Eq.\,(\ref{eq:assumeeqs015}). This assumption is applied in the following sections.

\begin{figure}[h]
\centering\subfigure[{The energy of the gravitational wave.}] {\includegraphics[scale=0.9,keepaspectratio]{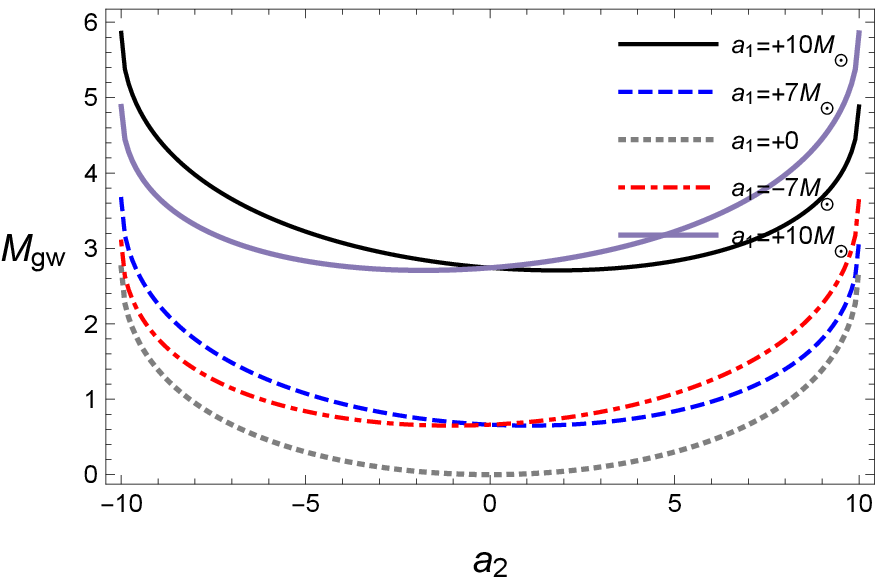}} \quad \centering\subfigure[{The ratio of the gravitational wave energy.}] {\includegraphics[scale=0.9,keepaspectratio]{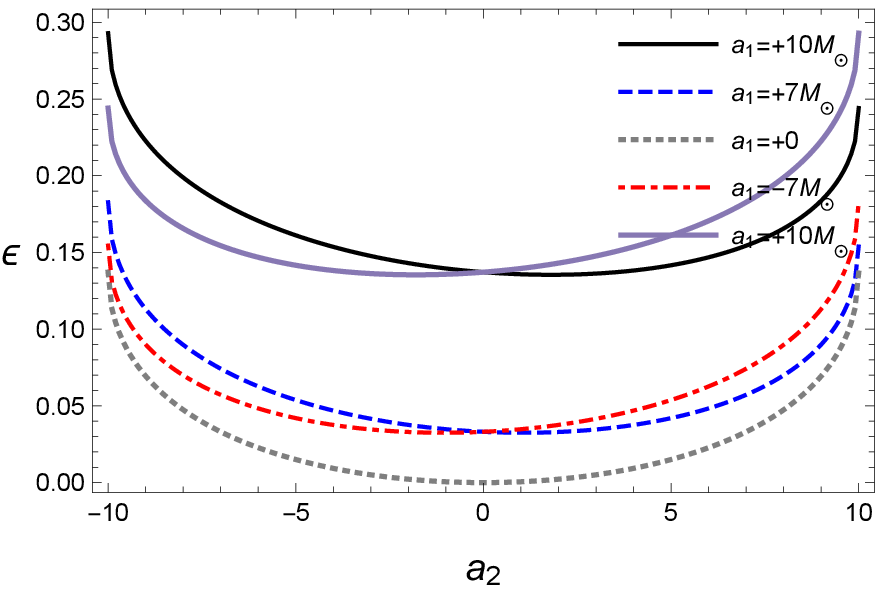}}
\caption{{\small The energy of the gravitational wave about $a_2$ for $M_1=10M_\odot$, $M_2=10M_\odot$, $\psi=0$, and $L_\text{orb}=0$.}}
\label{fig:fig1}
\end{figure}

We now investigate the energy of the gravitational wave with respect to the alignments of the spin angular momenta during coalescence. The effects of the spin parameters in the initial state are shown in Fig.\,\ref{fig:fig1} with respect to the second black hole's spin parameter $a_2$ for a given value of the first black hole's spin parameter.
The alignments with rotating axes are dependent on the sign of $a_1 a_2$ in Fig.\,\ref{fig:fig1}. For a plus sign, the two black holes are in a parallel alignment, and for a minus sign, they are in an antiparallel alignment. In Fig.\,\ref{fig:fig1}\,(a), the released energy $M_\text{gw}$ is the largest at the extremal values $a_2$ where the rotational energy of the secondary black hole is maximum because $M_\text{gw}$ is emitted from the reducible energy such as the rotational energy. Hence, the amount of energy $M_\text{gw}$ is proportional to the rotational energy in the initial state. For the alignments, the released energy is greater in an antiparallel alignment than in a parallel one owing to the contribution of the gravitational spin interaction, as expected in Eq.\,(\ref{eq:spininteraction012}). Owing to the dependence on the alignment, the minimum point of the released energy is located at the parallel alignment for a given rotational axis of the primary black hole. The ratio $\epsilon$ of the released energy with respect to the initial mass is shown in Fig.\,\ref{fig:fig1}\,(b). Interestingly, $0<\epsilon\leq 5\%$, which is similar to that of the LIGO observations of about $3\%$--$4.5\%$\cite{TheLIGOScientific:2016src,Abbott:2016nmj,Abbott:2017vtc,Abbott:2017gyy,Abbott:2017oio}. This implies that our approach based on the irreducible mass can provide results consistent with observations. Further, this supports the fact that most of the released energy originates from the reducible energy of the initial state of the binary system. Note that this is an improvement upon lower the upper limits given in previous studies\cite{Hawking:1971tu,Wald:1972sz,Gwak:2017zwm} to a realistic level.

\begin{figure}[h]
\centering
\subfigure[{$\psi-\epsilon$ diagram for $a_1=5M_\odot$.}] {\includegraphics[scale=0.9,keepaspectratio]{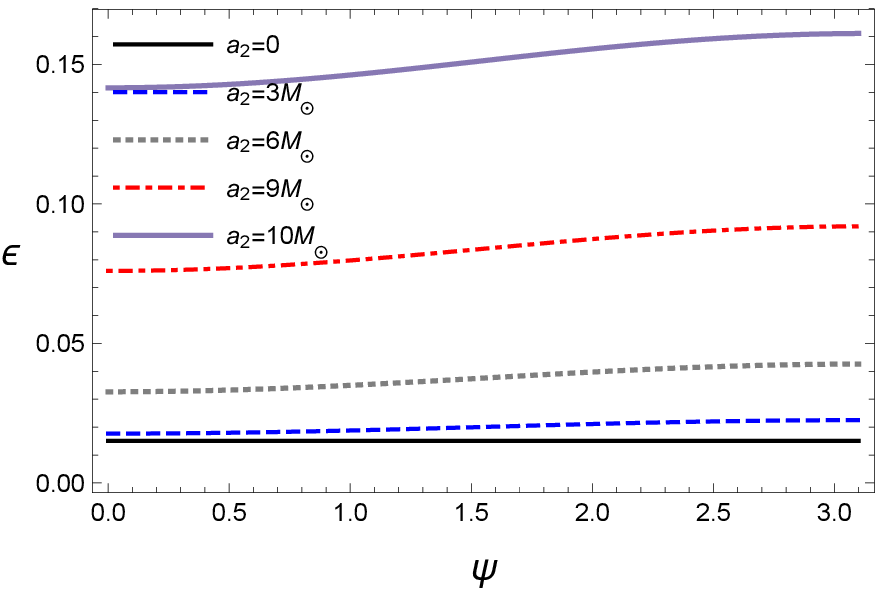}} \quad
\subfigure[{$\psi-\epsilon$ diagram for $a_1=10M_\odot$.}] {\includegraphics[scale=0.9,keepaspectratio]{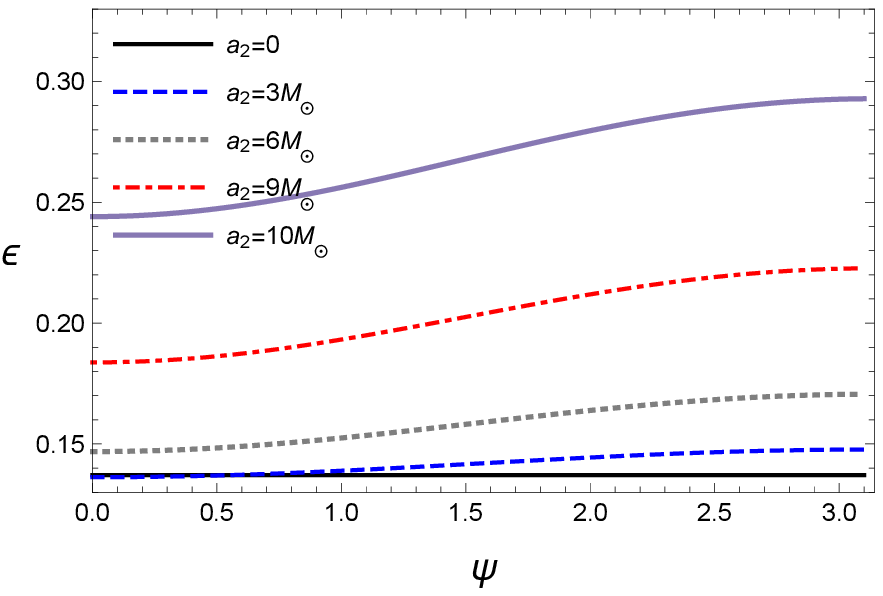}}
\caption{{\small The energy of the gravitational wave with respect to $\psi$ for $M_1=10M_\odot$, $M_2=10M_\odot$, and $L_\text{orb}=0$.}}
\label{fig:fig2}
\end{figure}

The detailed effects of the alignment can be obtained from the alignment angle $\psi$ in Eq.\,(\ref{eq:alignmentangle015}). The alignment angle $\psi$ shows the angle difference between the rotating axes of the primary and secondary black holes; thus, the parallel alignment corresponds to $\psi=0$, and the antiparallel alignment is $\psi=\pi$. The energy released by the gravitational wave with respect to $\psi$ is shown in Fig.\,\ref{fig:fig2}.
The energy released by the gravitational wave is the smallest at $\psi=0$ and increases as $\psi$ increases. Then, it attains a maximum at $\psi=\pi$. Hence, the attraction or repulsion due to the spin interaction plays an important role in the coalescence process. Further, the effects of the interaction are shown to be dependent on the angle $\psi$. The ratio $\epsilon$ of the gravitational wave energy also depends on the rotational energy for a given initial state. Although the primary black hole in Fig.\,\ref{fig:fig2}\,(a) has a spin parameter $a_1$ that is two times larger than that in Fig.\,\ref{fig:fig2}\,(b), the ratio $\epsilon$ in Fig.\,\ref{fig:fig2}\,(b) are greater than those in Fig.\,\ref{fig:fig2}\,(a) because the primary black hole in Fig.\,\ref{fig:fig2}\,(b) is an extremal black hole, which has the maximum rotational energy for a given mass. Note that the rotational energy exponentially increases as the spin parameter approaches that of the extremal black hole. Hence, we can expect that more rotational energy included in the initial state in Fig.\,\ref{fig:fig2}\,(b) is released as a gravitational wave than that in (a).

\begin{figure}[h]
\centering
\subfigure[{$\epsilon_\text{gw,rot}$ and $\epsilon_\text{f,rot}$ for $a_1=5M_\odot$.}] {\includegraphics[scale=0.9,keepaspectratio]{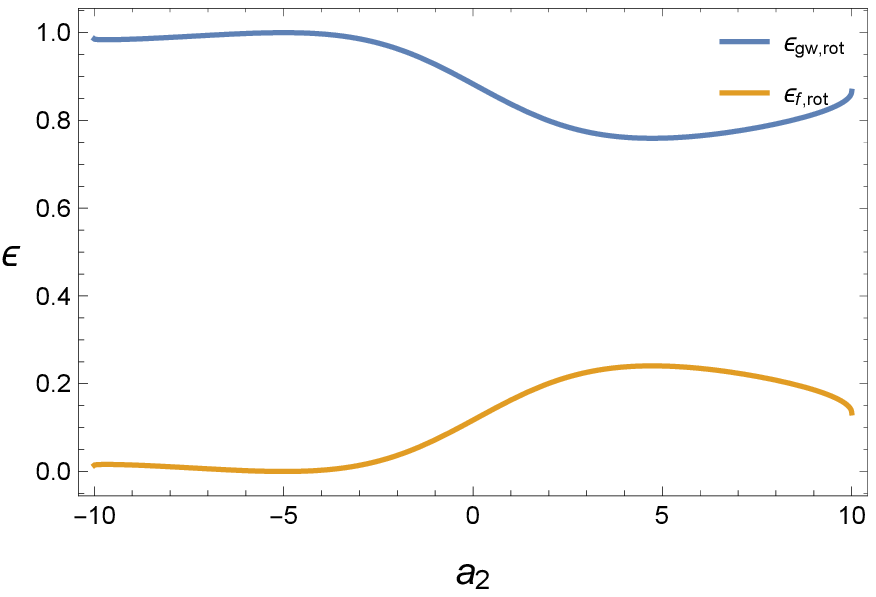}} \quad
\centering
\subfigure[{$\epsilon_\text{gw,rot}$ and $\epsilon_\text{f,rot}$ for $a_1=10M_\odot$.}] {\includegraphics[scale=0.9,keepaspectratio]{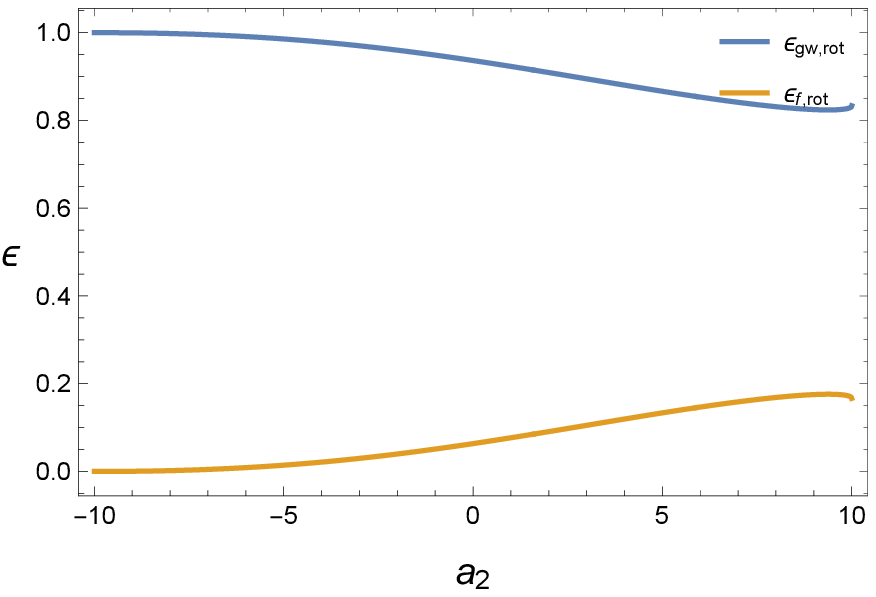}}\\
\caption{{\small $\epsilon_\text{gw,rot}$ and $\epsilon_\text{f,rot}$ with respect to $a_2$ for $M_1=10M_\odot$, $M_2=10M_\odot$, $\psi=0$, and $L_\text{orb}=0$.}}
\label{fig:fig4}
\end{figure}

Since the energy of the gravitational wave is assumed to be released from the reducible energy such as the rotational energy, we investigate the ratio of the energy of the gravitational wave to the rotational energies of the initial state and final black hole. The ratios of the energy of the gravitational wave and final black hole's rotational energy with respect to the initial masses are defined as
\begin{align}
\epsilon_\text{gw,rot}=\frac{M_\text{gw}}{M_\text{1,rot}+M_\text{2,rot}},\quad \epsilon_\text{f,rot}=\frac{M_\text{f,rot}}{M_\text{1,rot}+M_\text{2,rot}},
\end{align}
where the rotational energy is defined as $M_\text{rot}=M-M_\text{ir}$. Then, $\epsilon_\text{gw,rot}$ indicates the rotational energy released by the gravitational wave with respect to the initial rotational energy, and $\epsilon_\text{f,rot}$ is the remaining rotational energy in the final black hole. The interaction potential still plays an important role in releasing the gravitational wave; thus, the antiparallel alignment emits more energy than the parallel alignment, as shown in Fig.\,\ref{fig:fig4}.
In addition, owing to the conservation of the total energy, $\epsilon_\text{gw,rot}+\epsilon_\text{f,rot}=1$; thus, $\epsilon_\text{gw,rot}$ and $\epsilon_\text{f,rot}$ moves opposite to each other. Compared with Fig.\,\ref{fig:fig1}, the energy of the gravitational wave in Fig.\,\ref{fig:fig4}\,(a) is not maximized at the extremal value of the spin parameter $a_2$. Hence, the ratio related to the rotational energy is not exactly coincident with that of the total mass $\epsilon$. Further, in Fig.\,\ref{fig:fig1}, the maximum value of $\epsilon_\text{gw,rot}$ is located at $M_1 a_1 + M_2 a_2 \approx 0$, where the final black hole is almost close to a Schwarzschild black hole with a zero angular momentum because most of the rotational energy is released by the gravitational wave. As a special case, when the first black hole is extremal, if the second black hole is also extremal and antiparallel, the ratio becomes maximum, as shown in Fig.\,\ref{fig:fig4}\,(b). Here, most values of $\epsilon_\text{gw,rot}$ are quite large; therefore, we can expect that the final black hole is slowly rotating. However, the final black hole is at $a_\text{f}/M_\text{f}\approx 0.7$ in the LIGO observations. Thus, we need more angular momentum to fill this rotational energy gap.

\begin{figure}[h]
\centering\subfigure[{$M_2-M_\text{gw}$ diagram.}] {\includegraphics[scale=0.90,keepaspectratio]{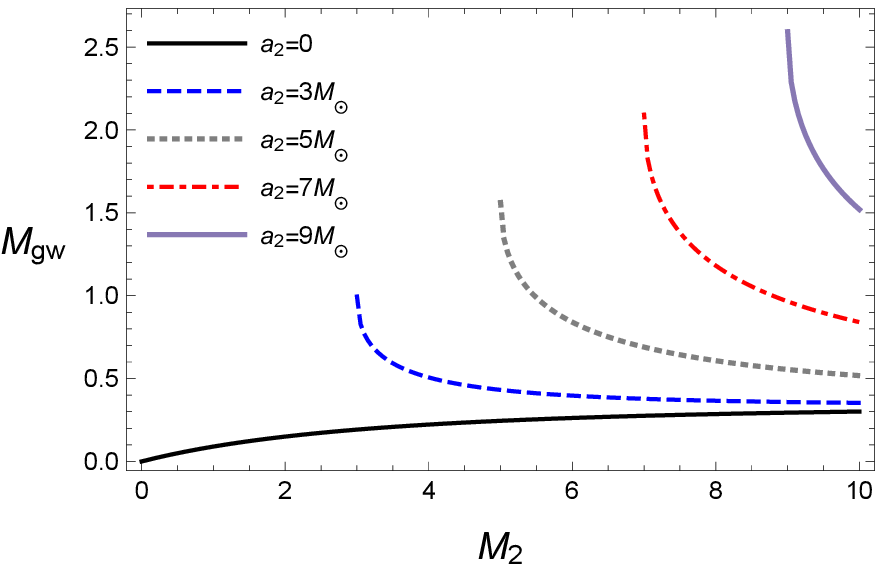}} \quad \centering\subfigure[{$M_2-\epsilon$ diagram.}] {\includegraphics[scale=0.9,keepaspectratio]{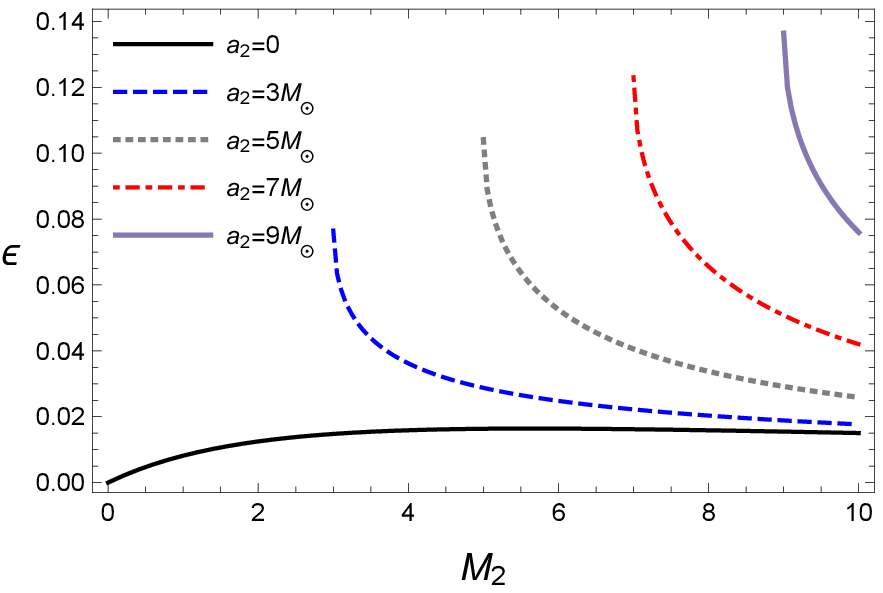}}
\caption{{\small The energy of the gravitational wave for $M_1=10M_\odot$, $M_2=10M_\odot$, $a_1=5M_\odot$, $\psi=0$, and $L_\text{orb}=0$.}}
\label{fig:f3}
\end{figure}

The masses of the initial states simply affect the energy released by the gravitational wave, as shown in Fig.\,\ref{fig:f3}. For a given mass of the primary black hole $M_1$, the released energy increases as $a_2$ increases in Fig.\,\ref{fig:f3}\,(a).
This is expected because a greater spin parameter for a given mass has more rotational energy in the initial state. The ratio $\epsilon$ also behaves in the same way, as shown in Fig.\,\ref{fig:f3}\,(b). Only the case of $a_2=0$ increases as $M_2$ increases, where the secondary black hole is a Schwarzschild black hole; thus, the increase in the secondary black hole's mass induces more rotational energy from the primary black hole. 
\begin{figure}[h]
\centering\subfigure[{$M_2-M_\text{gw}$ diagram.}] {\includegraphics[scale=0.9,keepaspectratio]{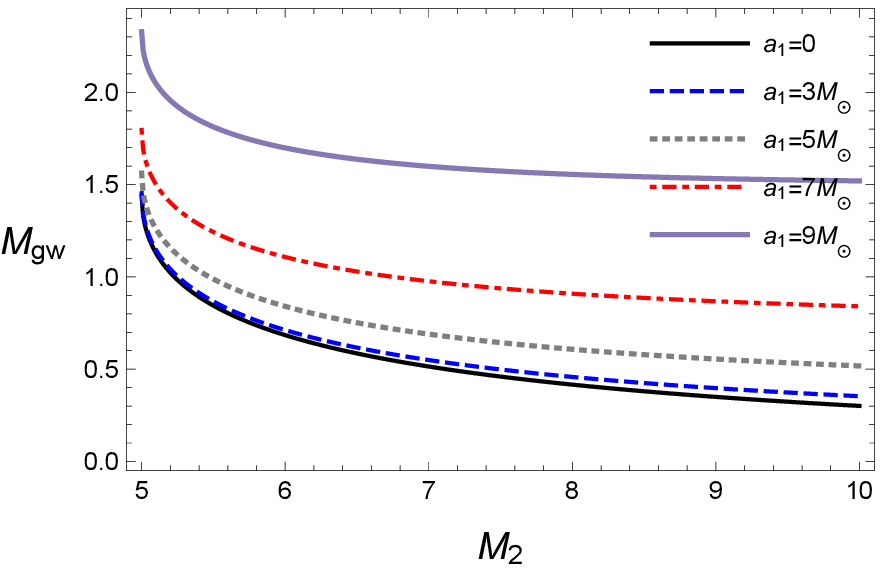}} \quad \centering\subfigure[{$M_2-\epsilon$ diagram.}] {\includegraphics[scale=0.9,keepaspectratio]{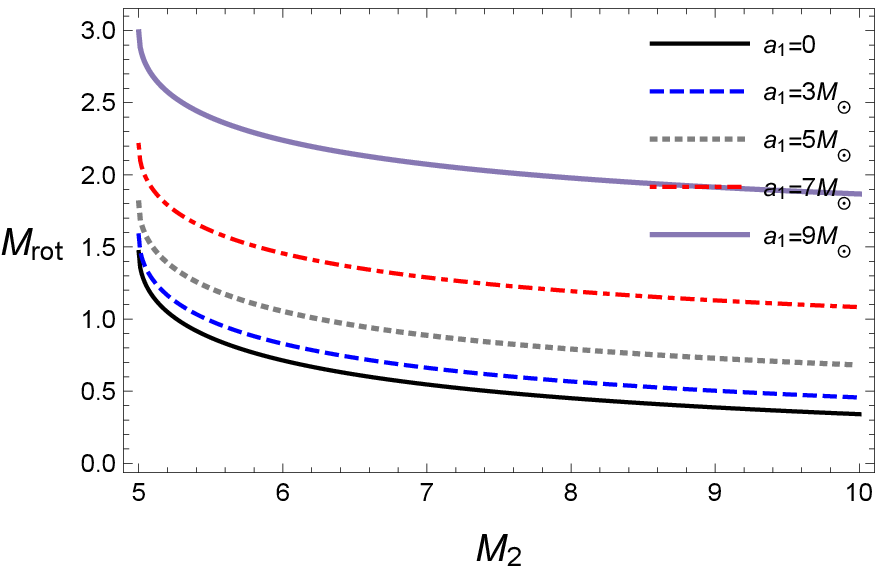}}
\caption{{\small The energy of the gravitational wave for $M_1=10M_\odot$, $M_2=10M_\odot$, $a_2=5M_\odot$, $\psi=0$, and $L_\text{orb}=0$.}}
\label{fig:f3b}
\end{figure}
The behaviors of the nonzero spin parameter cases are due to the amount of rotational energy in the initial state, as shown in Fig.\,\ref{fig:f3b}. For a secondary black hole with a fixed spin parameter, the energy released by the gravitational wave decreases as the mass of the secondary black hole increases in Fig.\,\ref{fig:f3b}\,(a) because the rotational energy of the secondary black hole decreases when the mass of the secondary black hole increases for a fixed spin parameter, as shown in Fig.\,\ref{fig:f3b}\,(b).

Here, we will investigate the effects of the orbital angular momentum $L_\text{orb}$ during the coalescence of the binary black hole. The orbital angular momentum causes spiral motions of the black holes before coalescence. Included in the total angular momentum of the black hole system, the orbital angular momentum similarly contributes as the spin angular momentum does, as shown in Eq.\,(\ref{eq:intpoten02}), in the particle limit. Owing to the orbital angular momentum, the angular momentum of the final black hole is expected to increase because most of the spin angular momenta cancel each other in the LIGO observations, where the black holes have antiparallel alignment.
\begin{figure}[h]
\centering\subfigure[{$L_\text{orb}-M_\text{f,rot}$ diagram for parallel alignments.}] {\includegraphics[scale=0.9,keepaspectratio]{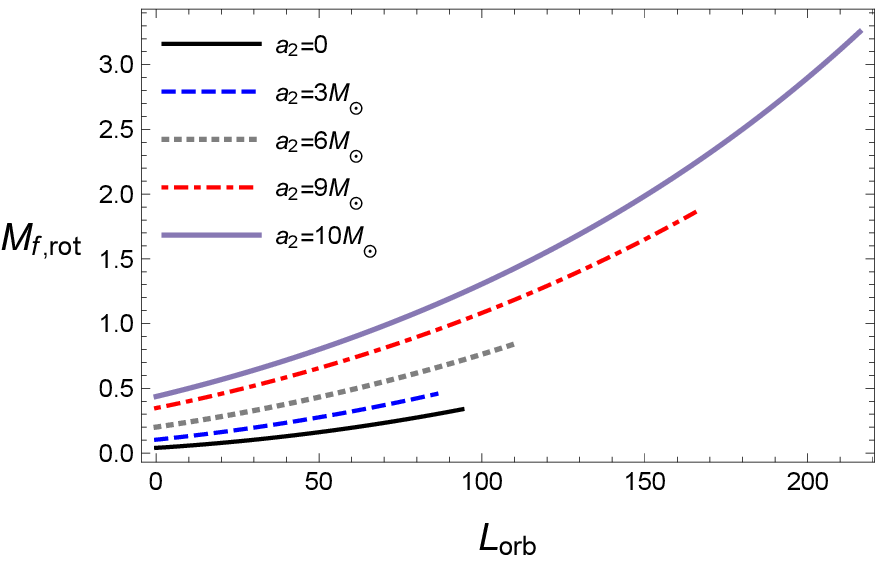}} \quad \centering\subfigure[{$L_\text{orb}-M_\text{f,rot}$ diagram for anti-parallel alignments.}] {\includegraphics[scale=0.9,keepaspectratio]{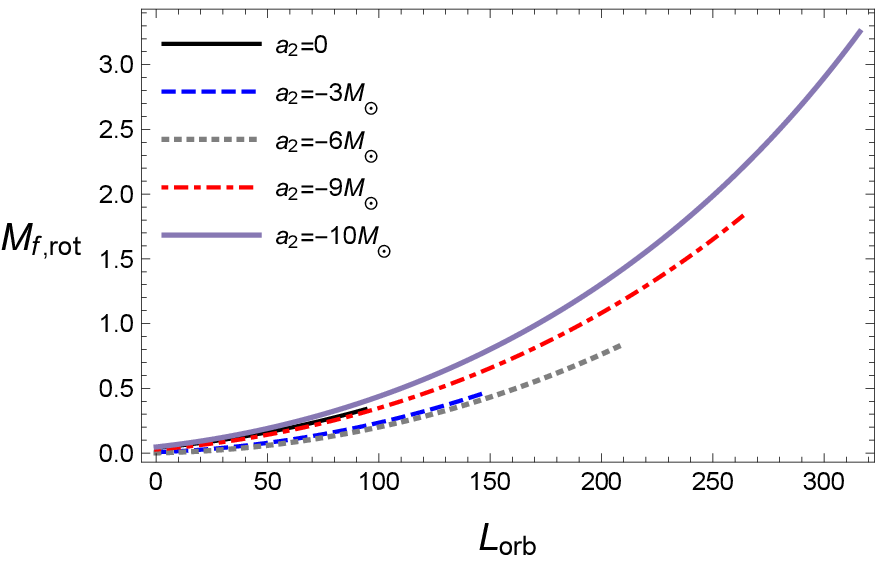}}\\
\caption{{\small The rotational energy of the final black hole for $M_1=10M_\odot$, $M_2=10M_\odot$, $a_1=5M_\odot$, and $\psi=0$.}}
\label{fig:fig5}
\end{figure}
Hence, the orbital angular momentum increases the rotational energy of the final black hole, as shown in Fig.\,\ref{fig:fig5}. For a given $L_\text{orb}$, the remaining rotational energy in the final black hole almost increases but is not exactly coincident because the released energy also depends on the alignment.
 \begin{figure}[h]
\centering\subfigure[{$L_\text{orb}-\epsilon$ diagram for parallel alignments.}] {\includegraphics[scale=0.9,keepaspectratio]{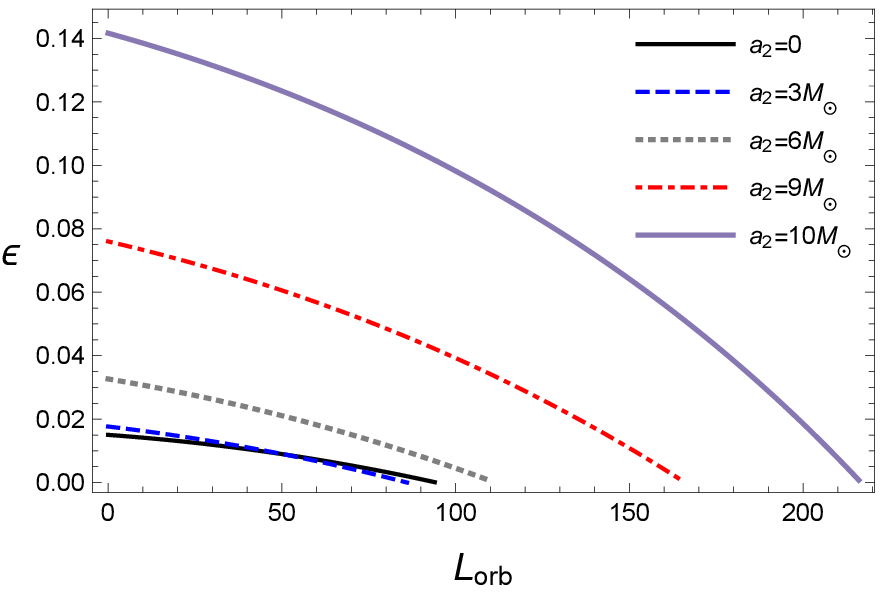}} \quad \centering\subfigure[{$L_\text{orb}-\epsilon$ diagram for anti-parallel alignments.}] {\includegraphics[scale=0.9,keepaspectratio]{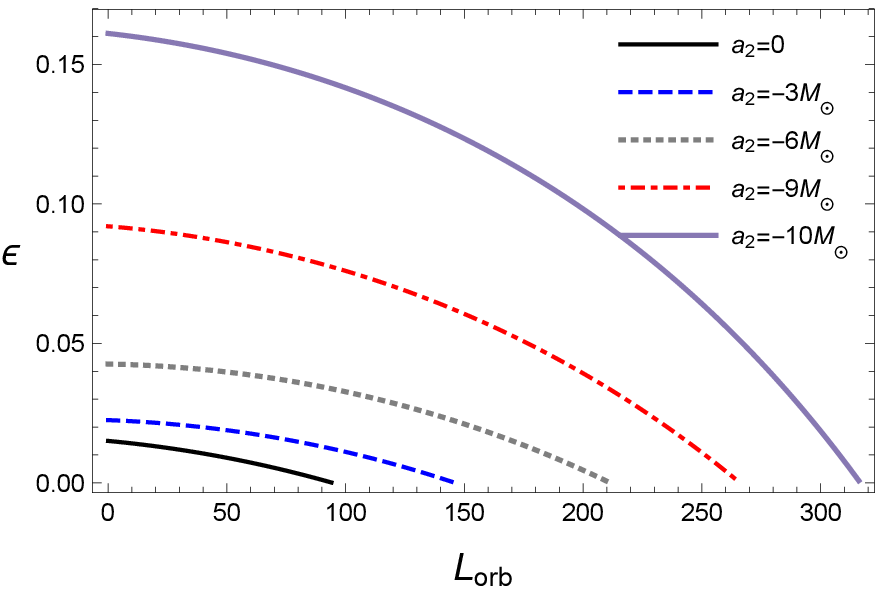}}\\
\caption{{\small The rotational energy of the final black hole for $M_1=10M_\odot$, $M_2=10M_\odot$, $a_1=5M_\odot$, and $\psi=0$.}}
\label{fig:fig5b}
\end{figure}
In Fig.\,\ref{fig:fig5}, for a primary black hole with $a_1=5M_\odot$, the orbital angular momentum of antiparallel cases is greater than those in the parallel cases. Further, there exist end points for a given initial condition in Fig.\,\ref{fig:fig5} due to the extremal condition of the final black hole. This is easily checked in terms of the ratio $\epsilon$ in Fig.\,\ref{fig:fig5b}. The orbital angular momentum $L_\text{orb}$ for a given initial state is proportional to the final spin parameter of the final black hole in Eqs.\,(\ref{eq:finalstate02}) and (\ref{eq:ratios01}). This saturates the final state of the extremal black hole; thus, to avoid saturation, more mass remains in the final black hole. This leads to a decrease in the ratio $\epsilon$. However, if a sufficiently large orbital angular momentum is set in the initial state, the total angular momentum is too large to form a final black hole. Thus, the end point $\epsilon=0$ implies that the final state is the extremal black hole, as shown in Fig.\,\ref{fig:fig5b}.

\begin{figure}[h]
\centering\subfigure[{$a_1-\epsilon$ diagram for $L_\text{orb}\geq0$.}] {\includegraphics[scale=0.9,keepaspectratio]{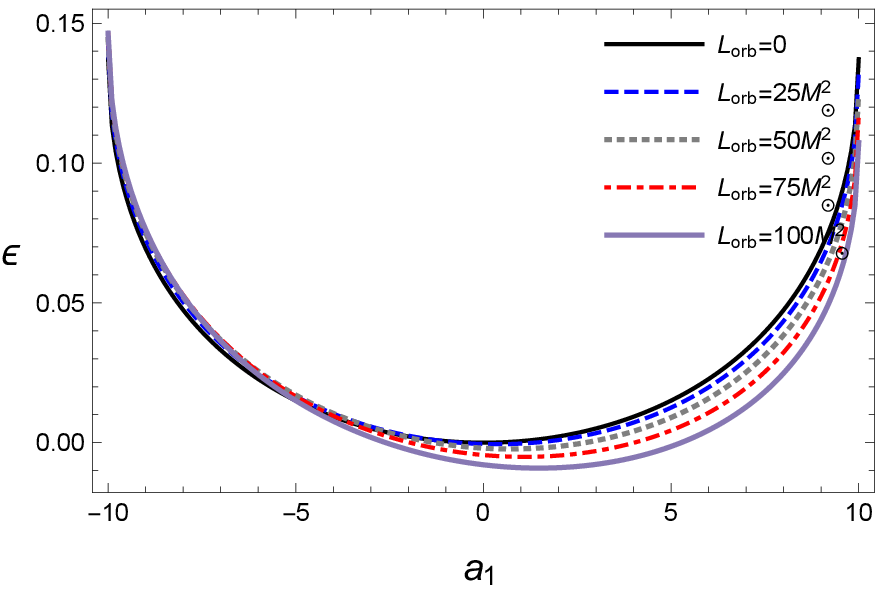}} \quad
\centering\subfigure[{$a_1-\epsilon$ diagram for $L_\text{orb}\leq0$.}] {\includegraphics[scale=0.9,keepaspectratio]{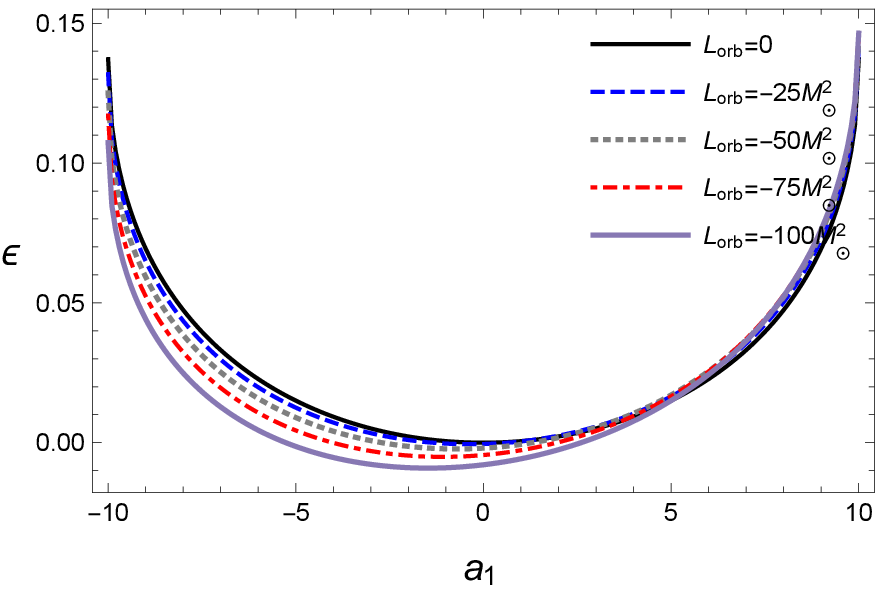}}
\caption{{\small The released ratio $\epsilon$ for $M_1=10M_\odot$, $M_2=10M_\odot$, $a_2=0$, and $\psi=0$.}}
\label{fig:fig7}
\end{figure}

The effects of the orbital angular momentum $L_\text{erb}$ with respect to the spin parameter $a_2$ contributes to the energy released by the gravitational wave, as shown in Fig.\,\ref{fig:fig7}.
According to the particle limit in Eqs.\,(\ref{eq:intpoten05}) and (\ref{eq:intpoten02}), there exists a gravitational spin--orbit interaction; thus, more energy is released by the gravitational wave in the antiparallel alignment than in the parallel one. The orbital angular momentum affects the energy of the gravitational wave with respect to the variation in the primary black hole's spin parameter $a_1$, as shown Fig.\,\ref{fig:fig7}.
\begin{figure}[h]
\centering\subfigure[{$a_2-\epsilon_\text{gw,rot}$ diagram for $a_1=6M_\odot$.}] {\includegraphics[scale=0.9,keepaspectratio]{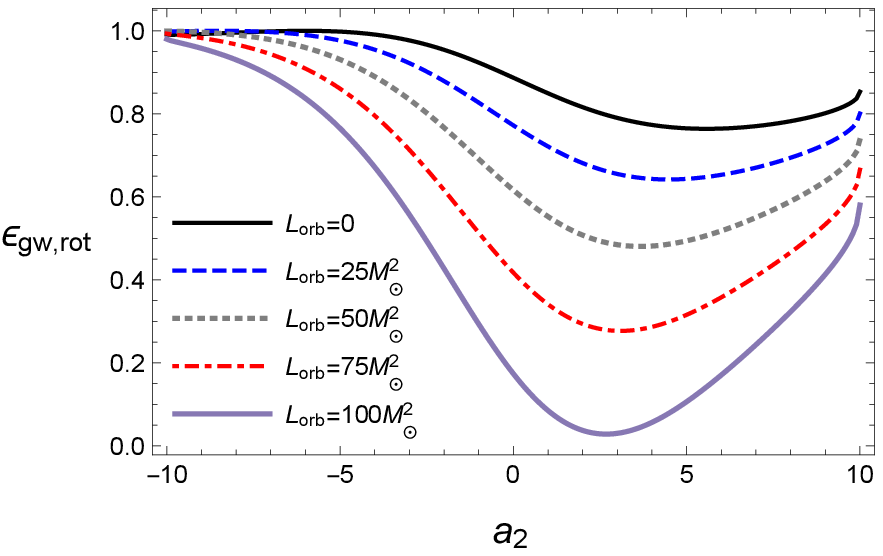}} \quad \centering\subfigure[{$a_2-\epsilon_\text{gw,rot}$ diagram for $a_1=9M_\odot$}] {\includegraphics[scale=0.9,keepaspectratio]{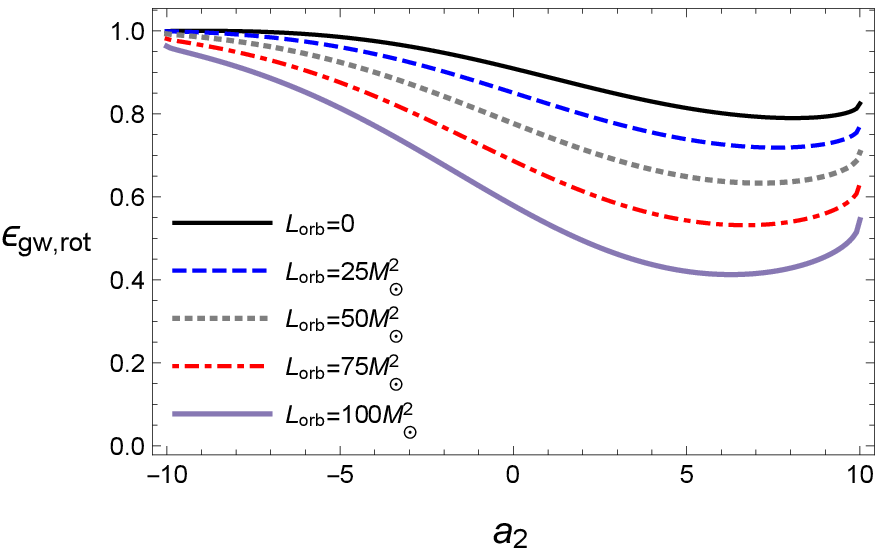}}
\caption{{\small The released ratio $\epsilon$ for $M_1=10M_\odot$, $M_2=10M_\odot$, and $\psi=0$.}}
\label{fig:fig8}
\end{figure}
In Fig.\,\ref{fig:fig7}, the orbital angular momentum moves the location of the minimum of the ratio $\epsilon$ to the parallel alignment. In addition, the ratio $\epsilon$ attains a maximum for an extremal black hole with an antiparallel alignment. Similarly, the rotational energy released by the gravitational wave is maximized at the extremal black hole, as shown in Fig.\,\ref{fig:fig8}. However, the ratio $\epsilon_\text{gw,rot}$ becomes small for a large value of the orbital angular momentum because the increase in $L_\text{orb}$ produces a final black hole with a large spin parameter $a_\text{f}$; thus, it becomes difficult to release the mass of the initial black holes by the gravitational wave to form the final black hole under the extremal bound. Here, the antiparallel alignment still releases more energy than the parallel alignment.

\section{Gravitational Wave Observations}\label{section5}

We have investigated contributions of various variables in our model of the coalescence of the binary black hole. Here, by combining all of the information about these variables, we will analyze the initial and final states of black holes including the orbital angular momentum in LIGO observations such as GW150914, GW151226, GW170104, GW170608, and GW170814\cite{TheLIGOScientific:2016src,Abbott:2016nmj,Abbott:2017vtc,Abbott:2017gyy,Abbott:2017oio}. In particular, we will mainly focus on the magnitude of the orbital angular momentum $L_\text{orb}$, which has not been thoroughly studied in previous studies by thermodynamics. In addition, the initial spin angular momenta, including the orbital angular momentum, will be estimated by applying our approach. By applying our approach to GW150914, we will introduce a general procedure for our investigation of the LIGO observations. Then, a similar analysis will be applied to other observations. Note that we will use source-frame masses related to detector-frame masses by applying the source redshift $z$.

\subsection{GW150914}

The source of GW150914 is a binary black hole merger with the effective inspiral spin parameter $\chi_\text{eff}=-0.07^{+0.16\pm0.01}_{-0.17\pm0.05}$, where the primary black hole is $M_1/M_\odot=35.8^{+5.3\pm0.9}_{-3.9\pm0.1}$ and $a_1/M_1=0.32^{+0.49\pm0.06}_{-0.29\pm0.01}$, and the secondary black hole is $M_2/M_\odot=29.1^{+3.8\pm0.1}_{-4.3\pm0.7}$ and $a_2/M_2=0.44^{+0.50\pm0.08}_{-0.40\pm0.02}$\cite{TheLIGOScientific:2016wfe}. Then, the coalescence of the binary black holes forms the final black hole having $M_\text{f}/M_\odot=62.0^{+4.1\pm0.7}_{-3.7\pm0.6}$ and $a_\text{f}/M_\text{f}=0.67^{+0.05\pm0.01}_{-0.07\pm0.02}$ with a released gravitational wave energy $M_\text{gw}/M_\odot=3.0^{+0.5}_{-0.5}$ of about $4.6\%$ of the total mass. Since the parameter ranges of $a_1$ and $a_2$ are much larger than that of $a_\text{f}$, we will find the proper parameter ranges of $a_1$ and $a_2$ by applying our approach. Finally, in agreement with the model, the orbital angular momentum and released angular momentum will be estimated. Here, our input values including errors are set to
\begin{align}
M_1/M_\odot=35.8^{+6.2}_{-4.0},\quad M_2/M_\odot=29.1^{+3.9}_{-5.0} \quad M_\text{f}/M_\odot=62.0^{+4.8}_{-4.3}, \quad a_\text{f}/M_\text{f}=0.67^{+0.06}_{-0.09}.
\end{align}
Then, the ranges of $a_1$, $a_2$, $L_\text{orb}$, $J_\text{tot}$, and $J_\text{gw}$ will be obtained with respect to the parameter range $\chi_\text{eff}=-0.07^{+0.17}_{-0.22}$.

We now utilize these observation data in our model. The basic framework is the same as Eqs.\,(\ref{eq:increaseirreduciblemass})--\,(\ref{eq:gwupper01}) in Sec.\,\ref{section3}. Since the spin parameters of the initial black holes are dependent on the effective inspiral parameter $\chi_\text{eff}$, our approach should be modified to add this parameter. In our definitions of the parameters, the effective inspiral parameter is given as
\begin{align}\label{eq:chidefine01}
\chi_\text{eff}= \left(\frac{M_1 \left(\frac{\vec{a}_1}{M_1}\right)+M_2 \left(\frac{\vec{a}_2}{M_2}\right)}{M_1+M_2}\right)\cdot \hat{L}_\text{orb} =\frac{ a_1 + a_2}{M_1+M_2},
\end{align} 
where the vector direction of the orbital angular momentum is fixed and $|\hat{L}_\text{orb}|=+1$. Further, we assume in Eq.\,(\ref{eq:chidefine01}) that the axes of the spin angular momenta are already aligned parallel or antiparallel to the axis of the orbital angular momentum. Hence, a positive spin parameter implies that the spin and orbital angular momenta rotate in the same direction, and a negative one in the opposite direction. In addition, we already assume the conservation of the irreducible mass. Then, from Eq.\,({\ref{eq:increaseirreduciblemass}}),
\begin{align}\label{eq:chidefine03}
\sqrt{r_1^2+a_1^2}+\sqrt{r_2^2+a_2^2}=\sqrt{r_\text{f}^2+a_\text{f}^2}.
\end{align}
In combination with Eqs.\,(\ref{eq:chidefine01}) and (\ref{eq:chidefine03}), the magnitudes and alignments of $a_1$ and $a_2$ can be obtained. Then, the magnitude of the orbital angular momentum can be calculated from the ratio of the gravitational wave $\epsilon\approx \epsilon_\text{M}\approx \epsilon_\text{J}$, which is rewritten as 
\begin{align}\label{eq:ratiomj010}
\epsilon=\frac{M_\text{gw}}{M_1+M_2}=\frac{J_\text{gw}}{M_1 a_1 +M_2 a_2 +L_\text{orb}}.
\end{align}
From Eq.\,(\ref{eq:ratiomj010}), the orbital angular momentum becomes 
\begin{align}
L_\text{orb}=(M_1+M_2)a_\text{f}-M_1 a_1 - M_2 a_2.
\end{align}
This implies that the absence of the total angular momentum between the initial and final states with the released one is supplied from the initial orbital angular momentum.
\begin{figure}[h]
\centering\subfigure[{Spin parameters of initial black holes.}] {\includegraphics[scale=0.9,keepaspectratio]{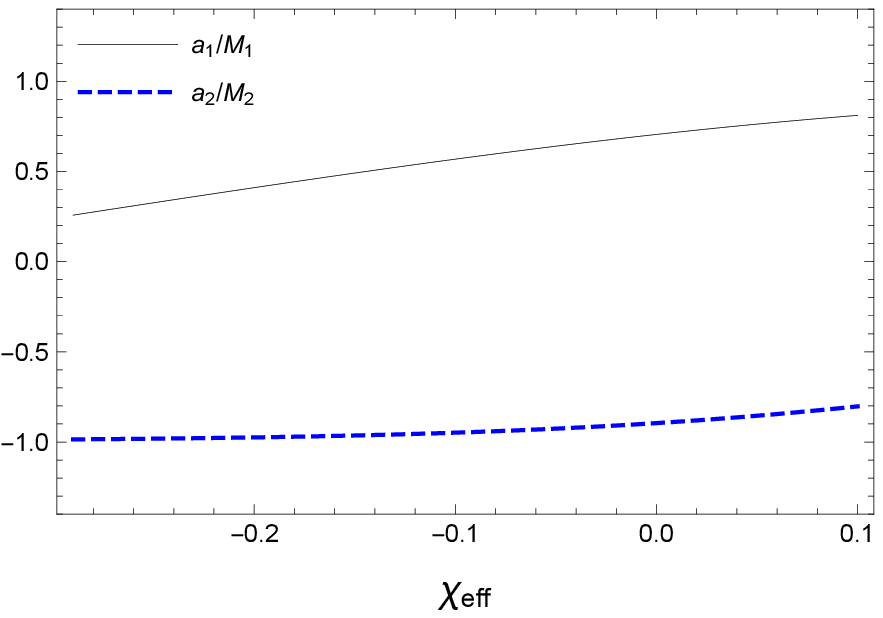}} \quad \centering\subfigure[{Orbital parameter of the initial binary black hole.}] {\includegraphics[scale=0.9,keepaspectratio]{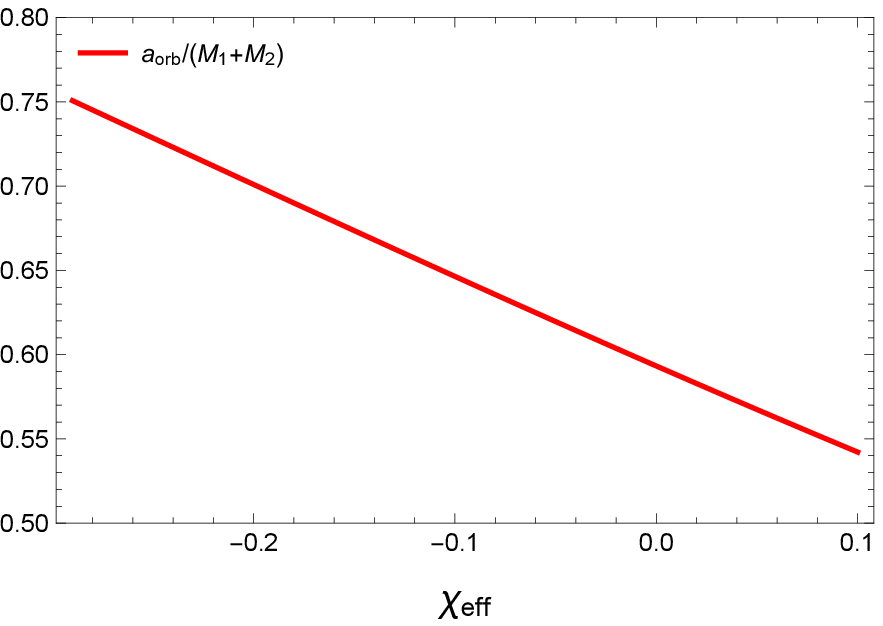}}
\caption{{\small Spin and orbital parameter with respect to $\chi_\text{eff}$ for $M_1=36.3M_\odot$, $M_2=28.6M_\odot$, $M_\text{f}=62.0M_\odot$, and $a_\text{f}/M_\text{f}=0.67$  with $a_1\geq 0$.}}
\label{fig:fig20a}
\end{figure}
This is supported from the observations of GW150914. i) There exists a difference between the sum of the spin angular momenta of the primary and secondary black holes and the final black hole. Approximately,
\begin{align}
M_\text{f}a_\text{f}-(M_1 a_1+M_2 a_2)\simeq (62M_\odot)^2\cdot 0.67-((35.8M_\odot)^2 \cdot 0.32+(29.1M_\odot)^2\cdot 0.44)\simeq +1800 M_\odot^2,
\end{align}
where an additional angular momentum of at least $1800M_\odot^2$ is needed for the final black hole to satisfy the conservation of the angular momentum. ii) Since the median value of $\chi_\text{eff}\simeq -0.07$ is around zero, the initial spin angular momenta of the initial black holes almost cancel each other; thus, the contribution of the spin angular momenta is very limited. Then, the spin angular momentum of the final black hole has to be supplied from other angular momenta in the initial state. Except the spin angular momenta, the only remaining angular momentum is the initial orbital angular momentum of the binary black hole system.
\begin{figure}[h]
\centering\subfigure[{Spin and orbital parameters of primary and secondary black holes.}] {\includegraphics[scale=0.9,keepaspectratio]{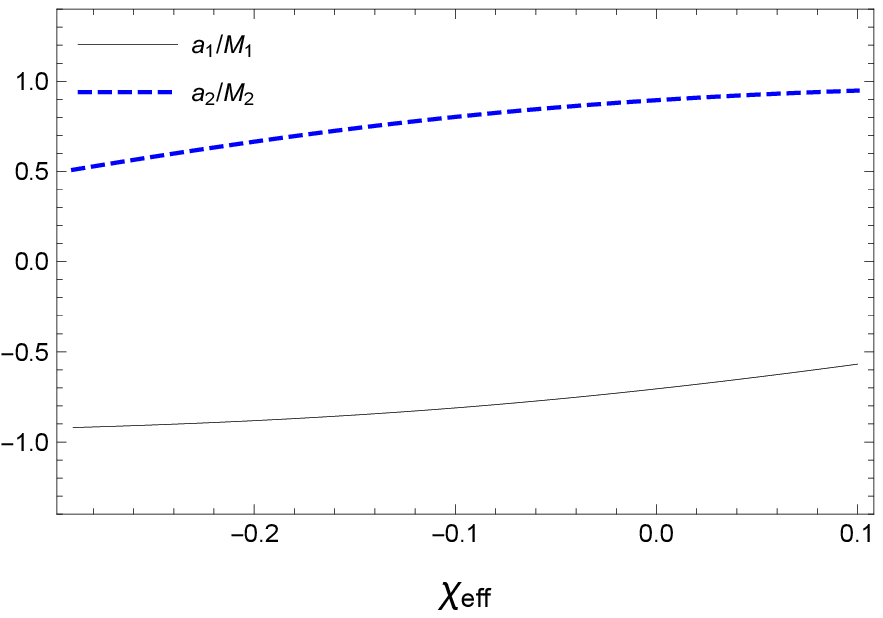}} \quad \centering\subfigure[{Orbital parameter of the initial binary black hole.}] {\includegraphics[scale=0.9,keepaspectratio]{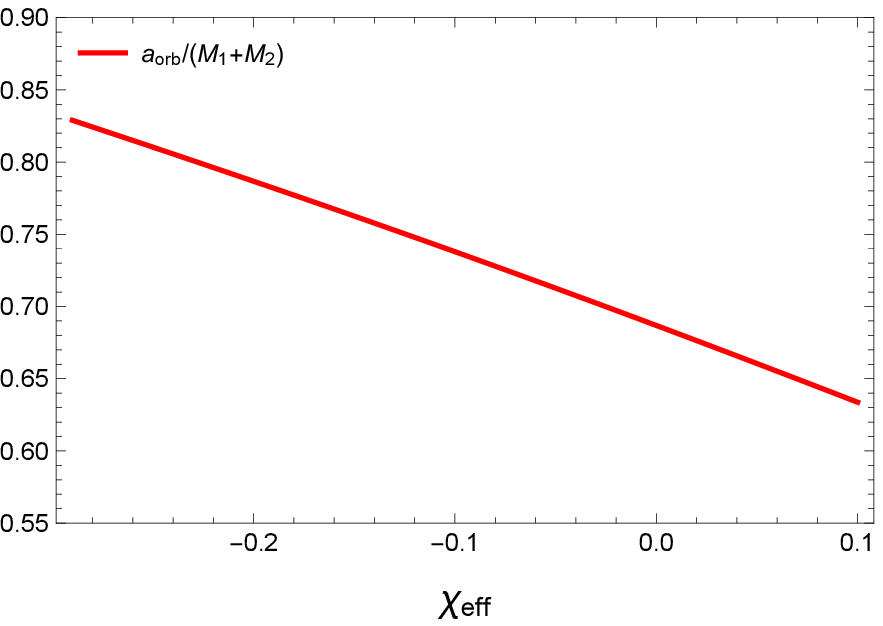}}
\caption{{\small Spin parameter with respect to $\chi_\text{eff}$ for $M_1=36.3M_\odot$, $M_2=28.6M_\odot$, $M_\text{f}=62.0M_\odot$, and $a_\text{f}/M_\text{f}=0.67$ with $a_1<0$.}}
\label{fig:fig20b}
\end{figure}
Satisfying Eqs.\,(\ref{eq:chidefine01}), (\ref{eq:chidefine03}), and (\ref{eq:ratiomj010}), the initial angular momenta can be divided into two cases, $a_1\geq 0$ and $a_1 <0$, as shown in Figs.\,\ref{fig:fig20a} and \ref{fig:fig20b}. Because $\chi_\text{eff}\simeq 0$, the spin parameters of the initial black holes are in antiparallel alignment in Figs.\,\ref{fig:fig20a}\,(a) and \ref{fig:fig20b}\,(a); therefore, the orbital angular momentum is of similar magnitude as the spin angular momentum of the final black hole in Fig.\,\ref{fig:fig20a}\,(b). The orbital angular momentum is given as a parameter with respect to the total mass of the initial state; therefore, $a_\text{orb}=\frac{L_\text{orb}}{(M_1+M_2)}$, as shown in Figs.\,\ref{fig:fig20a}\,(b) and \ref{fig:fig20b}\,(b). In consideration of ranges of input values, we can obtain magnitudes of estimated values of
\begin{align}
\frac{a_1}{M_1}=0.61_{-0.52}^{+0.39},\,\, \frac{a_2}{M_2}=0.93_{-0.74}^{+0.07},\,\, \frac{a_\text{orb}}{M_1+M_2}=0.64_{-0.28}^{+0.33},\,\, \frac{J_\text{tot}}{M_\odot^2}=2700_{-800}^{+1000},\,\, \frac{J_\text{gw}}{M_\odot^2}=120_{-110}^{+540},
\end{align}
where the median values are at $\chi_\text{eff}=-0.07$. The estimated spin parameters of the initial black holes are slightly larger than those of \cite{TheLIGOScientific:2016wfe} but within their error ranges. Further, as predicted from our approach, the ranges of spin parameters are tighter than those in \cite{TheLIGOScientific:2016wfe} and ensure the conservation of the angular momentum. Here, we newly estimate the sum of the total angular momentum in the initial state $J_\text{tot}$, which is a very large value compared with the initial spin angular momenta. This is based on the fact that the initial spin angular momentum is almost canceled owing to the alignment of the initial black holes. Hence, most of the spin angular momentum of the final state including the gravitational wave is provided from the orbital angular momentum of the binary black hole. This could be one reason why the frequency of the gravitational wave is almost proportional to the orbital frequency of the binary black hole system before coalescence. Thus, the orbital angular momentum is crucial in the coalescence of the binary black hole and the released gravitational wave.

We will repeat the same procedure for other observations of gravitational waves in the following subsections.

\subsection{GW151226}

The source of GW151226 is a binary black hole having masses of $M_1/M_\odot=14.2^{+8.3}_{-3.7}$ and $M_2/M_\odot=7.5^{+2.3}_{-2.3}$ with an inspiral spin parameter $\chi_\text{eff}=0.21^{+0.20}_{-0.10}$\cite{Abbott:2016nmj,TheLIGOScientific:2016pea}. The binary black hole system is only one example that has a positive value of $\chi_\text{eff}$\cite{Qin:2018vaa}. The coalescence of the binary black hole forms the final black hole whose mass and spin parameter are $M_\text{f}/M_\odot=20.8^{+6.1}_{-1.7}$ and ${a_\text{f}}/{M_\text{f}}=0.74^{+0.06}_{-0.06}$, respectively, with a released gravitational wave energy $M_\text{gw}/M_\odot=1.0^{+0.1}_{-0.2}$ of about $4.6\%$ of the total mass. By using these initial parameters, we can estimate that
\begin{align}
\frac{a_1}{M_1}=0.80_{-0.80}^{+0.20},\,\, \frac{a_2}{M_2}=0.91_{-0.78}^{+0.09},\,\, \frac{a_\text{orb}}{M_1+M_2}=0.48_{-0.32}^{+0.32},\,\, \frac{J_\text{tot}}{M_\odot^2}=330_{-80}^{+370},\,\, \frac{J_\text{gw}}{M_\odot^2}=14_{-14}^{+106},
\end{align}
where the orbital parameter is small compared with the initial spin parameters because the inspiral spin parameter is positive.

\subsection{GW170104}

The coalescence of the binary black hole having masses $M_1/M_\odot=31.2^{+8.4}_{-6.0}$ and $M_2/M_\odot=19.4^{+5.3}_{-5.9}$ forms the final black hole whose mass and spin parameter are $M_\text{f}/M_\odot=48.7^{+5.7}_{-4.6}$ and ${a_\text{f}}/{M_\text{f}}=0.64^{+0.09}_{-0.20}$, respectively\cite{Abbott:2017vtc}. The inspiral spin parameter of the binary black hole is estimated as $\chi_\text{eff}=-0.12^{+0.21}_{-0.30}$, which also includes zero in its range. This coalescence releases energy in terms of a gravitational wave, as much as $M_\text{gw}/M_\odot=2.0^{+0.6}_{-0.7}$, which is approximately $4.0\%$ of the initial total mass of the binary black hole. For these parameter ranges, we can obtain the magnitudes of the initial spin parameters as
\begin{align}
\frac{a_1}{M_1}=0.41_{-0.41}^{+0.59},\,\, \frac{a_2}{M_2}=0.97_{-0.96}^{+0.03},\,\, \frac{a_\text{orb}}{M_1+M_2}=0.60_{-0.38}^{+0.52},\,\, \frac{J_\text{tot}}{M_\odot^2}=1600_{-740}^{+1000},\,\, \frac{J_\text{gw}}{M_\odot^2}=59_{-55}^{+366}.
\end{align}

\subsection{GW170608}

GW170608 is released from the binary black hole merger whose component masses are $M_1/M_\odot=12^{+7}_{-2}$ and $M_2/M_\odot=7^{+2}_{-2}$ with the inspiral spin parameter $\chi=0.07^{+0.23}_{-0.09}$\cite{Abbott:2017gyy}. The coalescence of the binary black hole produces the final black hole having the mass $M_\text{f}/M_\odot=18.0^{+4.8}_{-0.9}$ and spin parameter $a_\text{f}/M_\text{f}=0.69^{+0.04}_{-0.05}$. The energy of GW170608 is about $M_\text{gw}/M_\odot=0.85^{+0.07}_{-0.17}$, which is approximately $4.5\%$ of the total mass of the system. Using our model, the estimated initial angular parameters are obtained as
\begin{align}
\frac{a_1}{M_1}=0.68_{-0.67}^{+0.32},\,\, \frac{a_2}{M_2}=0.97_{-0.63}^{+0.03},\,\, \frac{a_\text{orb}}{M_1+M_2}=0.52_{-0.32}^{+0.34},\,\, \frac{J_\text{tot}}{M_\odot^2}=240_{-50}^{+190},\,\, \frac{J_\text{gw}}{M_\odot^2}=12_{-10}^{+67}.
\end{align}

\subsection{GW170814}

The source of GW170814 is a binary black hole coalescence. The binary black hole has two components having masses $M_1/M_\odot=30.5^{+5.7}_{-3.0}$ and $M_2/M_\odot=25.3^{+2.8}_{-4.2}$ with the effective inspiral spin parameter $\chi_\text{eff}=0.06^{+0.12}_{-0.12}$\cite{Abbott:2017oio}. The coalescence of the binary black hole produces the final black hole whose mass and spin parameter are $M_\text{f}/M_\odot=53.2^{+3.2}_{-2.5}$ and $a_\text{f}/M_\text{f}=0.70^{+0.07}_{-0.05}$, respectively. The emitted energy of the gravitational wave is $M_\text{gw}/M_\odot=2.7^{+0.4}_{-0.3}$, which is about $4.6\%$ of the total mass of the system. Then, spin parameters consistent with the observations are estimated to be
\begin{align}
\frac{a_1}{M_1}=0.81_{-0.54}^{+0.09},\,\, \frac{a_2}{M_2}=0.85_{-0.39}^{+0.15},\,\, \frac{a_\text{orb}}{M_1+M_2}=0.60_{-0.24}^{+0.29},\,\, \frac{J_\text{tot}}{M_\odot^2}=2100_{-400}^{+700},\,\, \frac{J_\text{gw}}{M_\odot^2}=97_{-95}^{+413}.
\end{align}

\section{Summary}\label{sec6}

We have investigated the coalescence of a binary black hole with a released gravitational wave by constructing a model using Kerr black holes with an orbital angular momentum. In particular, this construction is expected to provide a more detailed analysis of the spin and orbital angular momenta of the binary system. Located far from each other in the initial state, two Kerr black holes having orbital angular momenta slowly come together to form the final Kerr black hole.

In the basic framework, we apply three assumptions in our approach: i) the conservation of energy, ii) the conservation of angular momentum, and iii) the conservation of the irreducible mass. Since we consider the irreducible mass, the mass of the Kerr black hole is divided into irreducible and reducible masses. Since our model should be coincident with particle absorption for a Kerr black hole, we obtain the constraints $\epsilon_\text{M}\approx\epsilon_\text{J}\approx\epsilon$ and identify that the irreducible mass can be approximately conserved for slowly moving black holes. Owing to the conservation of the irreducible mass, the upper limit of the energy released by a gravitational wave is very close to the actual energy of the gravitational wave; hence, we assume that $M_\text{gw,upper} \approx M_\text{gw}$. Further, analytical descriptions of the gravitational spin--orbit and spin--spin interactions are obtained for a variation of our model. Interestingly, these analytical forms of $U_\text{orb,int}$ and $U_\text{spin,int}$ exactly correspond to the results from particle absorption and the MPD equations.

Under the constraints and from an analytical analysis of particle absorption and the MPD equations, we have numerically applied our approach for the coalescence of binary black holes having equal masses. In this case, the range of energies of the released gravitational wave includes about $3.0\%$--$4.5\%$. This is very important advantage of our model because most of the LIGO observations support that the released energy ratio of the gravitational wave is about $4.5\%$. Therefore, we could provide A more realistic analysis using our approach based on a simple thermodynamic description. The energy released by the gravitational wave depends on the alignments of the spin and orbital angular momenta in the initial state. Since the dependency of the alignment is the same those in the gravitational spin--orbit and spin--spin interactions, as we expected, the released energy in an antiparallel alignment is greater than that in a parallel alignment for a fixed one of angular momenta. In addition, owing to form the final black hole, the orbital angular momentum is limited and has the maximum value.

Finally, we apply our approach to five LIGO observations of binary black hole mergers. We have estimated the parameters of the initial state using other parameters having small error ranges. Since the inspiral spin parameters are around zero in the most of observations, the large values of the spin angular momenta cancel each other. Hence, the spin angular momenta of the final black hole and gravitational wave have to be provided from the orbital angular momentum, which becomes a large value compared with the spin angular momentum. Using our approach, we have obtained $a_1$, $a_2$, $a_\text{orb}$, $J_\text{tot}$, and $J_\text{gw}$.

We have shown that the binary black hole merger can be approximated as the coalescence of two Kerr black holes having an orbital angular momentum in consideration of the irreducible mass. This implies that the energy source of the gravitational wave is the reducible energy such as the rotational energy and kinetic energy included the mass of the Kerr black hole.

\vspace{10pt} 

{\bf Acknowledgments}

This work was supported by the National Research Foundation of Korea (NRF) grant funded by the Korea government (MSIT) (NRF-2018R1C1B6004349).

\end{document}